\documentclass[11pt]{article}
\usepackage{amsmath,amssymb}
\usepackage{hyperref}

\renewcommand{\Im}{\operatorname{Im}}
\renewcommand{\Re}{\operatorname{Re}}

\newsavebox{\uuunit}
\sbox{\uuunit}
    {\setlength{\unitlength}{0.825em}
     \begin{picture}(0.6,0.7)
        \thinlines
        \put(0,0){\line(1,0){0.5}}
        \put(0.15,0){\line(0,1){0.7}}
        \put(0.35,0){\line(0,1){0.8}}
       \multiput(0.3,0.8)(-0.04,-0.02){10}{\rule{0.5pt}{0.5pt}}
     \end {picture}}

\newcommand{\ba}{\begin{eqnarray*}}
\newcommand{\ea}{\end{eqnarray*}}
\newcommand{\ban}{\begin{eqnarray}}
\newcommand{\ean}{\end{eqnarray}}

\newcommand{\cN}{{\cal N}}

\newcommand{\cS}{{\cal S}}

\newcommand{\cZ}{{\cal Z}}

\newcommand{\cD}{{\cal D}}

\newcommand{\sfQ}{{\mathsf{Q}}}

\newcommand{\sfB}{{\mathsf{B}}}

\newcommand{\im}{{\rm Im \,}}

\newcommand{\mbf}[1]{{\boldsymbol {#1} }}
\newcommand{\zed}{{\mathbb Z}} 
\newcommand{\real}{{\mathbb R}} 
\newcommand{\torus}{{\mathbb T}}

\def\e{{\,\rm e}\,}

\def\dd{{\rm d}}

\newcommand{\sign}{\mathrm{sgn}}

\def\beq{\begin{equation}}
\def\bee{\begin{equation}}
\def\eeq{\end{equation}}
\def\bea{\begin{eqnarray}}
\def\eea{\end{eqnarray}}
\def\bd{\begin{displaymath}}
\def\ed{\end{displaymath}}

\numberwithin{equation}{section}

\begin{document}

\thispagestyle{empty}
{}


\vskip -3mm
\begin{center}
{\bf\LARGE
\vskip - 1cm
Indefinite theta functions \\ [2mm]
and black hole partition functions}

\vspace{10mm}

{\large
{\bf Gabriel Lopes Cardoso$^+$,}
{\bf Michele Cirafici$^+$,}\\ \vskip 3mm
{\bf Rog\'erio Jorge$^{\times}$,}
{\bf Suresh Nampuri$^\dagger$}

\vspace{1cm}

{\it 
$^+$
Center for Mathematical Analysis, Geometry, and Dynamical Systems\\ [1mm] 
Departamento de Matem\'atica and LARSyS, 
Instituto Superior T\'ecnico
\\ [1mm]
1049-001 Lisboa, Portugal \\ [2mm] 
$^\times$
Instituto Superior T\'ecnico
\\ [1mm]
1049-001 Lisboa, Portugal \\ [2mm] 

$^\dagger$ 
Laboratoire de Physique Th\'eorique,  \'Ecole Normale Sup\'erieure \\ [1mm]
24 rue Lhomond, 75231 Paris Cedex 05, France 
}}

\vspace{5mm}

\end{center}
\vspace{5mm}

\begin{center}
{\bf ABSTRACT}\\
\end{center}

\noindent
We explore various aspects of supersymmetric black hole partition functions in 
four-dimensional toroidally
compactified heterotic string theory.  These functions suffer from divergences owing to the hyperbolic nature of the charge
lattice in this theory, which prevents them from having well-defined modular transformation properties.
In order to rectify this, we regularize these functions by converting the divergent series
into indefinite theta functions, thereby obtaining  fully regulated single-centered black hole partitions functions.

\clearpage
\setcounter{page}{1}

\tableofcontents

\section{Introduction and motivation}

 In string theory compactifications, certain classes of microscopic states in the Hilbert space of bound systems of solitons and strings can be described by black hole solutions at strong t'Hooft coupling. 
 In this context, exact counting functions have been developed that provide a statistical mechanical count of BPS states \cite{Dijkgraaf:1996it,Dabholkar:2004yr,Shih:2005uc,Jatkar:2005bh,David:2006ud,Sen:2008ta,Dabholkar:2008zy}.

For a class of string theory compactifications such as type II on Calabi-Yau threefolds $CY_3$, a topological twist 
creates a topological theory that captures the BPS aspects of the parent type II theory. 
In order to be able to write down the complete non-perturbative partition function of this theory, black holes must feed into the non-perturbative sectors of this theory and hence, writing down a well-defined black hole partition function becomes a significant step in this endeavour.

In this paper we explore aspects
of single-center black hole partition functions. We do this  in four-dimensional ${\cal N}=4$ compactifications,
since exact microstate counting formulae exist in these theories from which one can extract
black hole degeneracies.  In particular, we look at  four-dimensional toroidally
compactified heterotic string theory \cite{Dijkgraaf:1996it,Shih:2005uc}.

The first step 
in this program is to choose an ensemble to write down the single-center black hole partition function which can be used to extract the macroscopic black hole free energy.  We 
consider the mixed statistical ensemble first
introduced by OSV in \cite{Ooguri:2004zv}.  It can be motivated
by looking at partition functions in the near-horizon $AdS_3$ geometry
of  certain types of supersymmetric black holes.  These black hole partition functions were explored by
\cite{Shih:2005he,LopesCardoso:2006bg}, where they were computed using formal Poisson resummation techniques.
However,  these partition functions are divergent due to 
the indefinite nature of the charge lattice in the theory.  After an examination of the role played by the terms that contribute to the divergence in counting  single-centered black holes, we propose a regularization of the 
divergent series by converting the sums into
indefinite theta functions  following a prescription by Zwegers \cite{Zwegers:2008}, thereby obtaining  fully regulated black hole partitions functions
with well defined modular transformation properties.
As a guiding principle we demand that the leading contribution to the free energy of this
partition function equals the macroscopic black hole free energy.

We now summarize some of the salient features of the dyonic degeneracy formula and set up relevant notation for the discussions that follow in this paper.

\subsection{Notation and background material}

Upon compactification of the type II string on $K3 \times \torus^2$ physical charges are valued in the lattice $\Gamma^{6,22} \simeq H_2 (K3 ; \zed) \oplus 3 \, \Gamma^{1,1}$. Here $\Gamma^{1,1}$ is the hyperbolic lattice with bilinear form
\begin{equation}  
\Gamma^{1,1} = \left( \begin{matrix} 0 & 1 \\ 1 & 0 \end{matrix} \right) \;,
\end{equation}
while the intersection form of the homology lattice of $K3$ decomposes into $\Gamma^{3,19} = \oplus 2 \, \Gamma_{(- E_8)} \oplus 3 \, \Gamma^{1,1}$. The two vectors $\mbf Q$ and $\mbf P$ encoding the quantum numbers transform as a doublet under the S-duality group. The S-duality group is identified with the electric-magnetic duality in the heterotic frame and hence, these vectors can be labelled as electric and magnetic even if the individual charges are described in type II language. In this paper we will freely switch between the heterotic and type II dual frames, confident that no confusion should arise.

The T-duality invariant charge bilinears of the theory are the norm squares of the electric and
magnetic vectors, and their scalar product, explicitely $-Q = \mbf{Q} \cdot \mbf{Q}$, $-P = \mbf{P} \cdot \mbf{P}$ and $R = \mbf{Q} \cdot \mbf{P}$.
The degeneracies of a class of micro-states\footnote{We focus on $1/4$ BPS states with discrete invariant ${\rm gcd} ({\mbf Q} \wedge {\mbf P}) =1$ \cite{dabholkar:2007vk}.}
are expressed in terms of data associated with an auxiliary genus two Riemann surface. They are encoded in the expansion of the Siegel modular form
\begin{equation}
\frac{1}{\Phi_{10}(\sigma , \rho , v)}= \sum_{Q,P\,\in \,2\mathbb{Z},\, R\,\in\,\mathbb{Z}} d(Q,P,R) 
\, {\rm e}^{- \pi i (Q \, \sigma+ P \, \rho + R \, (2 \, v-1) )} \; ,
\label{eq:deg-sum}
\end{equation}
where the chemical potentials for the T-duality invariant bilinears parametrize the period matrix of said genus two Riemann surface
 \begin{equation}
  \left ( \begin{array}{cc} \rho & v\\v & \sigma \end{array}\right ) \ .
\label{2period}
\end{equation}
The degeneracies $d(Q, P, R)$ are non-vanishing for $Q \leq 2, P \leq 2$. For single-centered BPS black holes $Q$ and $P$ are negative while $QP - R^2 \gg 1$, and hence convergence of the $Q$ and $P$ sums in (\ref{eq:deg-sum}) requires that ${\rm Im} \, \rho = M_1 \gg 1$ and
${\rm Im} \,\sigma = M_2 \gg 1 $.  Invariance of \eqref{eq:deg-sum} under the large diffeomorphisms of the genus two Riemann surface given by
\begin{eqnarray}
\rho &\rightarrow& \rho + 1 \;, \nonumber\\
\sigma &\rightarrow& \sigma + 1 \;, \nonumber\\
v &\rightarrow& v + 1 \;,
\end{eqnarray}
ensures that we can always set the real parts of the chemical potentials to 
\begin{eqnarray}
0 \leq {\rm Re} \, \sigma < 1 \,, \, 0 \leq {\rm Re} \, \rho <1 \,,\, 0 \leq {\rm Re} \, v < 1\;.
\end{eqnarray}
In order to derive this counting formula, one typically chooses a canonical dyonic configuration where some of the individual charge quantum numbers are actually expressed in terms of invariant charge bilinears.
 For example, if one chooses
\begin{eqnarray}
{\mbf Q} &=& (q_0, -p^1,0,0,0, \dots, 0) \;, \nonumber\\
{\mbf P} &=&  (q_1,0, p^3,p^2,0, \dots, 0) \ ,
\end{eqnarray}
with $p^1 = 1$ and $p^2=1$, a simple computation using \eqref{charge-bilinears} gives
\begin{eqnarray} Q &=& 2 \, q_0 \ ,\nonumber\\ P &=& -2 \, p^3 \ ,\nonumber\\R &=& - q_1\ . 
\end{eqnarray}
Therefore equation \eqref{eq:deg-sum} can be rewritten by trading the sum over T-duality invariants with a sum over individual charges. The advantage of this rewriting is that it gives a direct path for comparing microscopic Hilbert space degeneracies to a macroscopic partition function over black hole backgrounds. The gravitational picture for these dyonic configurations includes extremal single-centered black holes which, at a specific point in their moduli space, have a near-horizon geometry described by a  BTZ black hole in $AdS_3$ \cite{Nampuri:2007gw}. Approaching the horizon of the BTZ black hole yields an $S^1$ fibration over $AdS_2$. The dual conformal field theory \cite{Maldacena:1997de} has a central charge defined by the charge $p^3$ which sets the scale for the $AdS_3$ space; all states in this CFT are labeled as excitations above the vacuum by the quantum number $q_0$ and the angular momentum $q_1 $ of the BTZ black hole.  This provides a macroscopic partition function which counts single-centered black hole attractor geometries in a statistical ensemble where the $p^I$ are held fixed and the $q_I$ are summed over\footnote{
In the $\cN=2$ theory, $I$ runs over $I=0,1,a$, where $a= 2, \dots, n$, with $n$ denoting the number of $\cN=2$ abelian vector multiplets coupled to $\cN=2$ supergravity.}.
In a general charge configuration, the fixed $p^I$ define the 
 $AdS_3$ spacetime, while the $q_I$ determine the BTZ excitations. Physically, this mixed ensemble captures all states in the near-horizon geometry of the black hole and should, in principle, capture the holographic entropy of the black hole, which is localized at the horizon. 

This mixed statistical ensemble was first introduced in \cite{Ooguri:2004zv} in the context of $\cN=2$ Calabi-Yau compactifications of type II string theory. Hence, we are motivated to 
write down a black hole
partition function in the mixed ensemble as  
\begin{equation}
\cZ_{\rm OSV} (p^I, \phi^I) = \sum_{q_I \in \Lambda_e} d(q_I, p^I) \, {\rm e}^{\pi \, q_I \phi^I } \;,
\label{osv-pf}
\end{equation}
where $\Lambda_e$ denotes the lattice of electric charges in the large volume polarization, the variables $\phi^I$ play the role of chemical potentials to be held fixed, and $d (q_I,p^I)$
denotes the absolute number (or a suitable index of) micro-states with electric/magnetic charges $(q_I, p^I)$.   
Observe that \eqref{osv-pf} is invariant under the shifts
\begin{eqnarray}
\phi^I \rightarrow \phi^I + 2 i \;.
\label{eq:shift-q-p}
\end{eqnarray}
This formal invariance is a consequence of the fact that the charges are quantized and integer valued.

In this paper we propose that the appropriate definition of the sum over the electric charges $q_a$
is in terms of indefinite theta functions. This will then also ensure the invariance under 
the shifts $\phi^a \rightarrow \phi^a + 2 i$. 
Definitions and properties of indefinite theta functions are briefly summarized in Appendix \ref{app-ind-theta-fun}.

\subsection{Models of interest}

In order to be able to use indefinite theta functions to define the sums over electric  charges, these have to belong to sub-lattices defined in terms of quadratic forms 
of signature $(r-1,1)$, respectively.
This is the case
in string models with
$\cN=2$ spacetime supersymmetry.  However, models for which exact microstate degeneracies of dyonic black holes
are known
are models with $\cN=4$ (or even $\cN=8$) supersymmetry.

In order to be able to apply the indefinite theta function regularization to $\cN=4$ models we will focus on a subset of $\cN=4$ charges,  which we denote by $(q_I, p^I)$ (with $I=0, 1, \dots, n$),
and we will
consider an effective $\cN=2$ description of these models 
based on prepotentials of the form
\begin{equation}
F^{(0)} (Y) = - \tfrac12 \, \frac{Y^1 \, Y^a \, C_{ab} \, Y^b}{Y^0} \;\;\;,\;\;\; a = 2, \dots, n \;,
\label{eq:heterotic-prep}
\end{equation}
where $n$ denotes the number of $\cN=2$
vector multiplets coupled to $\cN=2$ supergravity
and the symmetric matrix 
$C_{ab}$ appearing in \eqref{eq:heterotic-prep} has signature $(1,n-2)$, as required by the 
consistent coupling of vector multiplets to $\cN=2$ supergravity
\cite{Ferrara:1989py,Aspinwall:2000fd}.  
In this $\cN=2$ description, we take 
the associated charge bilinears and $C_{ab}$ to satisfy the same conditions as they do in $\cN=4$. We can think of these models as appropriate sub-sectors of the $\cN=4$ models where the charges associated with the extra indefinite directions have been set to zero  such that the rank of the intersection form $C_{ab}$ and the gauge group is $n-1$.
In this paper we focus on toroidally compactified heterotic string theory
\footnote{Hence  $n \leq 18$, since we are only keeping charges
associated  with $\oplus 2 \, \Gamma_{(- E_8)} \oplus 2 \, \Gamma^{1,1}$.}.  Accordingly, we consider
integer valued charges $(q_I, p^I)$ and integer valued matrices  
$C_{ab}$ and $C^{ab}$, so that $q_a C^{ab} q_b \in 2 \mathbb{Z}$, 
$p^a C_{ab} p^b \in  2 \mathbb{Z}$. 
The T-duality subgroup of the $\cN=4$ duality
group that operates on the charges $(q_I, p^I)$ is $SO(2, n-1)$.
The T-duality invariant charge bilinears 
are 
\begin{equation}
Q = 2 q_0 p^1 - q_a C^{ab} q_b \;\;\;,\;\;\;
P = - 2 p^0 q_1 - p^a C_{ab} p^b \;\;\;,\;\;\; R = p^0 q_0 - p^1 q_1 + p^a q_a \;.
\label{charge-bilinears}
\end{equation}
The heterotic moduli fields are denoted by $S = - i Y^1/Y^0$ (the heterotic dilaton-axion field) and 
$T^a = - i Y^a/Y^0$.  The effective $\cN=2$ description  also involves, in addition to $F^{(0)}$,
the S-duality invariant 
coupling function $F^{(1)}(S, \bar S)$.

As mentioned above we restrict ourselves to single-centered black hole states. These satisfy 
the following conditions, 
\begin{eqnarray}
S + \bar S > 0 \;, \, 
Q< 0 \;,\; P< 0 \;,\;  Q P - R^2 > 0 \;.
\end{eqnarray}
The black hole attractor mechanism relates the near horizon values of the moduli fields
$S$ and $T^a$ to the charges as \cite{Behrndt:1996jn}
\begin{eqnarray}
2 \sqrt{Q P - R^2} = |Y^0|^2 \, (S + \bar S) (T + \bar T)^a C_{ab} (T + \bar T)^b  =
\frac{\varrho_a \, 
C^{ab} \, \varrho_b}{|Y^0|^2 (S + \bar S)} \;, 
\label{entro-modu}
\end{eqnarray}
where 
\begin{equation}
\varrho_a = p^0 \, q_a + p^1 \, C_{ab} \, p^b \;.
\label{rho-vec}
\end{equation}
This implies ${\varrho_a \, 
C^{ab} \, \varrho_b} >0$.

Finally another ingredient we will need are indefinite theta functions. Aspects of the theory are reviewed in Appendix \ref{app-ind-theta-fun}. As in the case of ordinary theta functions, indefinite theta functions depend on a quadratic form $\sfQ: 
\mathbb{R}^r  \longrightarrow \mathbb{R}$ and its associated bilinear form $\sfB : \real^r \times \real^r \longrightarrow \real^r$. However in this case the quadratic form $\sfQ$ has signature $(r-1,1)$. To retain convergence and modularity, one weights the sum with additional factors $\rho$ as 
\begin{equation}
\vartheta (z; \tau) = \sum_{n \in \mathbb{Z}^r} \rho(n + a; \tau) \, {\rm e}^{ 2 \pi i \,  \tau \, \sfQ(n) + 2 \pi i \,  \sfB(n,z)} \;,
\label{ind-theta}
\end{equation}
where $\tau \in {\cal H}$ takes values in the complex upper half plane ${\cal H}$, and $z \in \mathbb{C}^r$, with $a, b \in \mathbb{R}^r$ defined by $z = a \, \tau + b$. The factor $\rho$ is the difference of two functions $\rho^c$, 
$\rho = \rho^{c_1} - \rho^{c_2}$. The $\rho^{c_i}$ depend on real vectors $c_i \in \mathbb{R}^r$ 
that satisfy $\sfQ(c_i) \leq 0$ and are given by \cite{Zwegers:2008}
\begin{equation}
\rho^c (n ; \tau) = \left\{ \begin{matrix} 
E \left (\frac{\sfB ( c , n)}{\sqrt{-\sfQ (c)}} \, \sqrt{\im \tau} \right) & \text{if} \, \sfQ (c) < 0 \\
\sign \left( \sfB (c, n ) \right) & \text{if} \, \sfQ (c) = 0
\end{matrix} \right.
\end{equation}
where $E$ and $\sign$ denote the error and sign function, respectively. In the main text we will take $\sfQ(c_i) <0$ for both $c_1$ and $c_2$.

The paper is organized as follows. We focus on the OSV ensemble and sum over the charges $q_0$ and $q_1$ 
without imposing any restrictions, 
following \cite{LopesCardoso:2006bg}.  
We compare the leading contribution to this sum with the macroscopic 
$1/4$ BPS single-center black hole free energy.
We then turn to the sum over the charges $q_a$ and specialize to the case $p^0 =0$ in order
to avoid a technical difficulty that arises when $p^0 \neq  0$. We regularize the sum over $q_a$ by converting it into an indefinite theta function.

\section{Evaluation of $\cZ_{\rm OSV}(p,\phi)$}\label{sec2}

In the following we consider the evaluation of the mixed partition function \eqref{osv-pf} in toroidally compactified heterotic string theory, using an effective $\cN=2$ description of this model based on \eqref{eq:heterotic-prep}.

\subsection{Summing over $(q_0, q_1)$}

We first 
sum over the charges $q_0$ and $q_1$ following 
\cite{Shih:2005he,LopesCardoso:2006bg}.
We convert the sum over $(q_0, q_1)$ into a sum over $(Q, P)$ using
\begin{eqnarray}
q_0 &=& \frac{1}{2 p^1}  \left( Q + \, q_a C^{ab} q_b \right) \,, \nonumber\\
q_1 &=& - \frac{1}{2 p^0}  \left( P + p^a C_{ab} p^b \right) \;,
\label{QPqq}
\end{eqnarray}
where, for the time being\footnote{We will later specialize to $p^0 =0$.},
we assume that both $p^0$ and $p^1$ are non-vanishing, i.e. $|p^0|\geq 1 \,,\, |p^1| \geq 1$.
In doing so, we need to ensure that when performing the sums over $Q$ and $P$,
we only keep those contributions that lead to integer-valued charges of $q_0$ and $q_1$.
These restrictions
can be implemented by inserting the series $L^{-1} \sum_{l =0}^{L-1} {\rm exp}[2 \pi i \, l \, K/L]$, where $K$
and  $L$ are integers (with $L$ positive), which projects onto all integer values for $K/L$.
The use of this formula leads to the following expression 
\begin{eqnarray}
\sum_{q_0, q_1} \,  d(q,p) \, 
{\rm e}^{\pi \, q_I \phi^I }
=\frac{1}{ |p^0 p^1|} \sum_{\tiny \begin{array}{c} l^0= 0 , \dots |p^1| -1\\ 
 l^1= 0 , \dots |p^0| -1 \end{array} }
L(R,\hat{\phi}^0,\hat{\phi}^1,\phi^a)\,
\label{eq:QPsum}
\end{eqnarray}
\,with $R$ given by
\begin{equation}
R = \frac{p^0}{2 p^1} \left(Q + q_a C^{ab} q_b \right) + \frac{p^1}{2 p^0} \left(P + p^a C_{ab} p^b \right)
+ q_a \, p^a \;,
\label{rel-R-charges}
\end{equation} 
and \cite{LopesCardoso:2006bg}
\begin{eqnarray}
 L(R,\hat{\phi}^0,\hat{\phi}^1,\phi^a)\,&=&\,\sum_{Q, P} \,  d(Q, P, R) \, \nonumber\\
&&{\rm exp}\left[ \frac{\pi \hat{\phi}^0}{2 p^1} \left(Q + q_a C^{ab} q_b \right) -
\frac{\pi \hat{\phi}^1}{2 p^0} \left(P  + p^a C_{ab}  p^b \right) 
+ \pi q_a \phi^a 
\right] \;,
\label{eq:Lsum}
\end{eqnarray}
where 
\begin{eqnarray}
\hat{\phi}^0 &=& \phi^0 + 2 i \, l^0 \;, \nonumber\\
\hat{\phi}^1 &=& \phi^1 + 2 i \, l^1 \;.
\end{eqnarray}
The range of the sums over $l^{0,1}$ enforces the condition that
only those summands, for which $(Q + q_a C^{ab} q_b)/2p^1$ and $(P + p^a C_{ab} p^b)/2 p^0$ are integers,
give a non-vanishing contribution to \eqref{eq:QPsum}.

Now we introduce an additional sum over a dummy variable $R'$ so as to be able to use the representation \eqref{eq:deg-sum}.
To this end, we use a complex variable $\theta = \theta_1 + i \theta_2$, and write
\begin{equation}
f(R) = \sum_{R'}  {\rm e}^{ - 2\pi \theta_2 (R - R')} \, f(R') \, {\rm e}^{\pi i (R' - R)}
\int_{0}^1 \dd \theta_1 \, {\rm e}^{ 2\pi i  \theta_1 (R - R')}  \;,
\end{equation}
which holds for integer valued $R, R'$. Here, $\theta_2$ is held fixed.
Then we introduce
\begin{eqnarray}
\sigma (\theta) &=& i \, \frac{\hat{\phi}^0}{2 p^1} - (2 \theta -1) \, \frac{p^0}{2p^1} \;, \nonumber\\
\rho (\theta) &=& - i \, \frac{\hat{\phi}^1}{2 p^0} - (2 \theta -1) \, \frac{p^1}{2p^0} \;, \nonumber\\
v (\theta) &=& \theta \;, \nonumber\\
{\tilde \phi}^a (\theta) &=& \phi^a + i \, (2 \theta -1) \, p^a \;.
\label{eq:srvthet}
\end{eqnarray}

Next, using \eqref{eq:deg-sum} and interchanging summations and integrations, we obtain \cite{LopesCardoso:2006bg}
\begin{eqnarray}
\sum_{q_0, q_1} d(q,p) \, {\rm e}^{\pi q_I \phi^I   } 
&=& \frac{1}{|p^0 p^1|}  \sum_{\tiny \begin{array}{c} l^0= 0 , \dots |p^1| -1\\ 
 l^1= 0 , \dots |p^0| -1 \end{array} }\,   \int_0^1 \dd \theta_1 
\, \frac{1}{\Phi_{10} (\sigma(\theta), \rho (\theta) , \theta)} \nonumber\\
&& \qquad {\rm exp} \left[ - i \pi \sigma(\theta)\, q_a C^{ab} q_b + \pi q_a {\tilde \phi}^a (\theta) - \pi i \, \rho (\theta) \, p^a C_{ab} p^b 
\right] \,.
\label{eq:Z-phi-theta}
\end{eqnarray}
We note that identifying the $d(q,p)$ on the lhs of \eqref{eq:Z-phi-theta} with the microcanonical dyonic degeneracy generated by the Siegel modular form $\,\frac{1}{\Phi_{10}}\,$, automatically fixes the arguments 
$(\sigma(\theta), \rho(\theta), \theta)$  
of the Siegel modular form to be the period matrix of a genus two Riemann surface, i.e. they have to take values in the Siegel upper half-plane
\begin{eqnarray}\label{eq:sig}
\Im \rho \;,\;  \Im \sigma& > &0 \,,\,\nonumber\\
\Im \rho \; \Im \sigma &> & (\Im \theta)^2 \,.
\end{eqnarray}
Applying these restrictions to \eqref{eq:srvthet} imposes the constraints,
\begin{eqnarray}\label{eq:phicond}
\frac{\phi^0-2 \theta_2 p^0}{2 p^1}\,&>&\,0\,,\nonumber\\
\frac{-\phi^1-2 \theta_2 p^1}{2 p^0}\,&>& \, 0\,,  \nonumber\\
\theta_2 \, \frac{( \phi^1 p^0-\phi^0 p^1)}{p^0 p^1} &>&  \frac{\phi^0 \phi^1}{2 p^0 p^1} \;.
\end{eqnarray}

Next, 
observe that \eqref{eq:Z-phi-theta}
 is invariant under the shifts \cite{Shih:2005he,LopesCardoso:2006bg}
\begin{equation}
\phi^0 \rightarrow \phi^0 + 2 i p^1 \, n \;\;\;,\;\;\;
\phi^1 \rightarrow \phi^1 + 2 i p^0 \, m \;,
\label{big-shift-01}
\end{equation}
with $n, m \in \mathbb{Z}$.  Namely, under these shifts, $\sigma(\theta)$ and 
$\rho(\theta)$ transform as 
\begin{equation}
\sigma(\theta) \rightarrow \sigma(\theta) - n \;\;\;,\;\;\;
\rho(\theta) \rightarrow \rho(\theta) + m \;,
\label{sig-rho-sh}
\end{equation}
and since $q_a C^{ab} q_b \in 2 \, \mathbb{Z} \;,\;  p^a C_{ab} p^b  \in 2 \, \mathbb{Z}$, the exponent in the integrand of 
\eqref{eq:Z-phi-theta} is invariant under the shifts \eqref{big-shift-01}. 
Using $\Phi_{10}(\sigma - n, \rho, v) = \Phi_{10}(\sigma, \rho, v)$ and
$\Phi_{10}(\sigma, \rho + m, v) = \Phi_{10}(\sigma, \rho, v)$, 
it follows that \eqref{eq:Z-phi-theta} is invariant under the shifts \eqref{big-shift-01}. This invariance, together with the sum over $l^0, l^1$-shifts, ensures that 
\eqref{eq:Z-phi-theta} is invariant under $\phi^0 \rightarrow \phi^0 + 2 i \,,\,
\phi^1 \rightarrow \phi^1 + 2 i$.  

$\Phi_{10}$ has various zeros \cite{Dijkgraaf:1996it}. The location of these zeros 
is parametrized in terms of five integers $m_1, n_1, m_2, n_2 \in \mathbb{Z}$, $j \in 2 \mathbb{Z} + 1$, which
are subject to the condition
\begin{equation}
m_1 \, n_1 + m_2 \, n_2 + \tfrac14 j^2 = \tfrac14 \; .
\label{rel-5int}
\end{equation}
The zeros are at
\begin{equation}
n_2 \, ( \rho \, \sigma - v^2) + j \, v + n_1 \, \sigma - m_1\,  \rho + m_2 =0 \;.
\label{location-zero}
\end{equation}
The zeros with $n_2$ encode the jumps in the degeneracies across walls of marginal stability corresponding to two centered small black holes which appear (or disappear) in the stable spectrum \cite{Sen:2007pg,Cheng:2007ch}.
The zeros with $n_2 \geq 1$
capture the entropy of single-center black holes \cite{Dijkgraaf:1996it}. 
The leading contribution to the entropy stems from the zeroes with $n_2 =1$.
Among them is the zero ${\cal D}$ with non-vanishing integers $n_2 = j =1$, i.e.
 ${\cal D} = 
\rho \sigma - v^2  + v= 0$.
In the following we focus on the zeros with $n_2 =1$. These zeros can be generated from 
the zero ${\cal D}$, which is described by $(m_1,n_1,m_2,n_2,j)=(0,0,0, 1, 1)$,
as follows.
First, observe that $\Phi_{10}(\sigma, \rho, v)$ is invariant under the discrete translations in $v$: $\Phi_{10}(\sigma, \rho, v + p) = \Phi_{10}(\sigma, \rho, v)$ with $p \in \mathbb{Z}$.  
Then, applying the shift transformation
$v \rightarrow v + p$ as well as \eqref{sig-rho-sh} to ${\cal D}$ results in zeros 
${\cal D}_{(n,m,p)}$
specified
by the integers $(n, m, -m n -p^2 +p , 1, 1-2p)$. In particular, ${\cal D}_{(0,0,0)} = {\cal D}$.
This provides a parametrization of the zeros 
with $(m_1, n_1, m_2, 1, j)$ satisfying \eqref{rel-5int}.
The same holds for the zeroes of $\Phi_{10}(\sigma(\theta), \rho(\theta), \theta)$, provided we make 
the compensating transformation $\phi^0 \rightarrow \phi^0 - 2 i  p \, p^0 \,,\, 
\phi^1 \rightarrow \phi^1 + 2 i p \, p^1$ when performing the shift $\theta \rightarrow \theta + p$. These compensating transformations constitute an invariance of \eqref{eq:Z-phi-theta}, as discussed above.

We may thus proceed as follows.
The integral \eqref{eq:Z-phi-theta} will be evaluated
in terms of the residues associated with the zeros of $\Phi_{10}$. 
Here we restrict ourselves to the zeros with $n_2 =1$ which, as we just discussed, can be parametrized in terms
of integers $n, m, p$. The contribution of these zeros can be accounted for by retaining the contribution of the zero
${\cal D}$
and extending 
the sum over $l^{0,1}$ to run over all positive and negative integers (i.e. $l^{0,1} \in \mathbb{Z}$) as well as
extending the range of integration to
$- \infty < \theta_1 < \infty$. 
Hence, the poles of $\frac{1}{\Phi_{10}}$  corresponding to $n_2 = 1 $ are characterized in terms of three numbers $m, n$ and $p$,  and by swapping the infinite ranges of $m$ and $n$ for the infinite ranges of $l^0$ and $l^1$, respectively, and swapping $p$ for the infinite range of $\theta_1$, we have fully expressed the entire subgroup of symmetries under which the poles corresponding to $n_2 = 1$ form a closed group, in terms of sums over two discrete variables and an integration over one continuous real variable.  

We now proceed with the evaluation of \eqref{eq:Z-phi-theta}, focussing on the
contribution of the zero ${\cal D}=0$.

\subsection{Free energy computation}

To compute the contribution from the zero ${\cal D}=0$, we follow the prescription given
in \cite{Cheng:2007ch}, as follows.  In the complex $\theta$-plane,
the contour of integration in 
\eqref{eq:Z-phi-theta} is
taken to be $- \infty <  \theta_1 < \infty$ (as we just discussed) with fixed $\theta_2$, either $\theta_2 > 0$ or $\theta_2 < 0$.
The $\theta$-dependent part of the exponential in \eqref{eq:Z-phi-theta} can be written as 
\begin{equation}
{\rm exp} \left[  \pi i  \,\frac{\theta}{p^0 p^1} \varrho_a \, 
C^{ac} \, \varrho_c \,
\right] \;,
\label{eq:expthet}
\end{equation}
with $\varrho_a$ given in \eqref{rho-vec}.
We consider single-center black holes so that  $\varrho_a C^{ab} \varrho_b > 0$.
 The choice of the sign of $\theta_2$ then depends on the sign of $p^0 p^1$. 
Namely, when $p^0 p^1 < 0$,  we take $\theta_2 > 0$.
We can then deform the contour to $\theta_2 \rightarrow - \infty$, where the integrand becomes vanishing.
In doing so, we pick up the contribution from the zero  ${\cal D}=0$, which will be specified below.
Here, the zero is encircled in a clockwise direction.  When 
$p^0 p^1 > 0$, we take $\theta_2 <0$. The contour can then be moved to  $\theta_2 \rightarrow \infty$, where the integrand is
again zero.  In doing so, we pick up the contribution from the zero  ${\cal D}=0$, but this time it is encircled
in a counter clockwise direction.  
Thus, 
we obtain a non-vanishing contribution provided we choose
the 
integration contour
to satisfy
$p^0 p^1 \,
\theta_2 <0$.  Then, the integral yields 
\begin{equation}
{\rm sgn} \left(p^0 p^1 \right) \, {\rm Res} \;,
\label{sign-sgn}
\end{equation}
where ${\rm Res}$ denotes the residue which we now compute.
Inserting \eqref{eq:srvthet} into the expression for 
${\cal D}= 
v + \rho \sigma - v^2 = 0$, we find
that the zero is located at the value (recall that now $l^{0,1} \in \mathbb{Z}$)
\begin{equation}
\theta_* = \frac12 - i \frac{ \hat{\phi}^0 \hat{\phi}^1 + p^1 p^0}{2(\hat{\phi}^0 p^1 - \hat{\phi}^1 p^0)} \;,
\label{eq:zero-Phi}
\end{equation}
which is complex, and hence away from the real $\theta_1$ axis.
In the vicinity of $\theta_*$, ${\cal D}$ takes the form
\begin{equation}
 {\cal D} (\theta) = 2 (\theta - \theta_*) \frac{(\hat{\phi}^0 p^1 - \hat{\phi}^1 p^0)}{4 i p^0 p^1} \;,
\end{equation}
while $\Phi_{10}$ takes the form 
$\Phi_{10} \approx {\cal D}^2 \, \Delta$ with 
\begin{eqnarray}
\Delta = \sigma^{-12} \, \eta^{24}(\gamma') \, \eta^{24}(\sigma') \;, 
\end{eqnarray}
where
\begin{equation}
\gamma' = \frac{\rho \sigma - v^2}{\sigma} \ , \qquad \sigma' = \frac{\rho \sigma - (v-1)^2}{\sigma} \;.
\end{equation}
For later convenience, we also introduce the notation
\begin{equation}
4 \pi \, \Omega (\theta) = - \ln \Delta (\theta) \;.
\end{equation}
Then, using \eqref{sign-sgn}, we obtain for  \eqref{eq:Z-phi-theta}
(we drop an overall numerical constant)
\begin{eqnarray}
\sum_{q_0, q_1} d(q,p) \, {\rm e}^{\pi q_I \phi^I   } 
&=& p^0 p^1 \sum_{l^0 \in \mathbb{Z}, l^1 \in \mathbb{Z}} \, 
\frac{1}{(\hat{\phi}^0 p^1 - \hat{\phi}^1 p^0)^2}
\nonumber\\
&& \hskip - 20mm
\frac{\dd}{\dd \theta} 
 {\rm exp} \left[ - i \pi \sigma(\theta)\, q_a C^{ab} q_b + \pi q_a {\tilde{\phi}}^a (\theta) - \pi i \, \rho (\theta) \, p^a C_{ab} p^b + 4 \pi \Omega(\theta)
\right]_{\theta = \theta_*}.
\label{eq:Z-phi-theta-res}
\end{eqnarray}
Using
\begin{eqnarray}
\frac{\dd \sigma(\theta) }{\dd \theta}{\Big|}_{\theta = \theta_*} &=& - \frac{p^0}{p^1} \;\;\;,\;\;\;
\frac{\dd \rho(\theta) }{\dd \theta}{\Big|}_{\theta = \theta_*} = - \frac{p^1}{p^0} \;, 
\end{eqnarray}
we obtain
\begin{eqnarray}
\label{result1q0q1}
\sum_{q_0, q_1} d(q,p) \, {\rm e}^{\pi q_I \phi^I   } 
&=& 
\sum_{l^0 \in \mathbb{Z}, l^1 \in \mathbb{Z}} \; 
\frac{M}{(\hat{\phi}^0 p^1 - \hat{\phi}^1 p^0)^2} \, \\
&& \hskip - 15mm
{\rm exp} \left[\pi \left(- i \sigma(\theta_*) q_a C^{ab} q_b - i \rho(\theta_*) p^a C_{ab} p^b + 
i (2\theta_*-1) q_a p^a  + q_a \phi^a + 4  \Omega(\theta_*)  \right) \right] \;,
\nonumber
\end{eqnarray}
where 
\begin{eqnarray}
M =  \left(p^0 q_a + p^1 C_{ab} p^b \right) C^{ac} \left(p^0 q_c + p^1 C_{cd} \, p^d \right) -  4 i \, p^0 p^1\, 
\frac{d \Omega (\theta)}{d \theta}\Big|_{\theta = \theta_*}  \;.
\label{eq:def-M}
\end{eqnarray}
Now, following \cite{LopesCardoso:2006bg}, 
we generalize the definitions of $S$ and $Y^0$ given in  \eqref{S-phi-p} and \eqref{eq:back-val-Y}
and 
introduce the shifted fields 
\begin{eqnarray}
S &=& \frac{- i \hat{\phi^1}+ p^1}{\hat{\phi}^0 + i p^0} = 
 \frac{- i \phi^1 + (p^1 + 2 l^1 )}{\phi^0 + i (p^0 + 2 l^0)} \;, \nonumber\\
\bar S &=& \frac{i \hat{\phi}^1 + p^1 }{\hat{\phi}^0 - i p^0} = 
\frac{i \phi^1 + (p^1 - 2 l^1)}{\phi^0 - i (p^0-2 l^0)} \;, \nonumber\\
Y^0 &=& \frac12 \left( \hat{\phi}^0 + i p^0 \right) = \frac12 \left( \phi^0 + i (p^0 + 2 l^0) \right) \;, \nonumber\\
{\bar Y}^0 &=& \frac12 \left( \hat{\phi}^0 - i p^0 \right) = \frac12 \left( \phi^0 - i (p^0 - 2 l^0) \right) \;.
\label{eq:SbSl}
\end{eqnarray}
Observe that in the presence of the $l^0, l^1$-shifts, $\bar S$ and $\bar{Y}^0$ are not any longer 
the complex conjugate of $S$ and of $Y^0$, respectively.
Using \eqref{eq:SbSl}, we obtain
\begin{eqnarray}
\sigma(\theta_*) &=&  
\frac{i}{S + \bar{S}} \;, \nonumber\\
\rho(\theta_*)&=& i \frac{S \bar{S}}{S + \bar{S}} \;, \nonumber\\
2 \theta_* - 1 &=& \frac{S -\bar{S}}{S + \bar{S}} \;, \label{theta2} \nonumber\\
(\hat{\phi}^0 p^1 - \hat{\phi}^1 p^0)^2 &=& 4 (S + \bar S)^2 \left(Y^0 \bar{Y}^0 \right)^2 \;,
\nonumber\\
\gamma'(\theta_*) &=& i S \;, \nonumber\\
\sigma'(\theta_*) &=& i \bar S \;, \nonumber\\
4 \pi \Omega(\theta_*) &=& 4 \pi \Omega(S, \bar S) = - 12 \ln ( S + \bar S) - \ln \eta^{24}(S) - \ln \eta^{24} (\bar S) 
\;,
\label{rel-theta-st}
\end{eqnarray}
as well as
\begin{eqnarray}
\frac{\dd \sigma'(\theta) }{\dd \theta}{\Big|}_{\theta = \theta_*} &=& - \frac{1}{p^0 p^1} \, (S + \bar S)^2 \left(Y^0 \right)^2 \;, \nonumber\\
\frac{\dd \gamma'(\theta) }{\dd \theta}{\Big|}_{\theta = \theta_*} &=& - \frac{1}{p^0 p^1} \, (S + \bar S)^2 \left({\bar Y}^0 \right)^2 \;.
\end{eqnarray}
Inserting these expressions into \eqref{result1q0q1}, we get (dropping again a numerical constant)
\begin{eqnarray}
\label{res-q01-sum}
\sum_{q_0, q_1} d(q,p) \, {\rm e}^{\pi q_I \phi^I   } 
&=& 
\sum_{l^0 \in \mathbb{Z}, l^1 \in \mathbb{Z}}  \; 
\frac{M}{(S + \bar S)^2 \left(Y^0 \bar{Y}^0 \right)^2} \, \\
&& \hskip - 25mm
{\rm exp} \left[
\frac{\pi}{S + \bar S} \left( q_a C^{ab} q_b + S {\bar S} p^a C_{ab} p^b + (S + \bar S) q_a \phi^a + 
i (S - \bar S) q_a p^a  \right) + 4 \pi \, \Omega(S, \bar S)
\right] \;,
\nonumber
\end{eqnarray}
with $M$ expressed as
\begin{eqnarray}
\label{eq:M-expl}
M &=&  
\left(p^0 q_a + p^1 C_{ab} p^b \right) C^{ac} \left(p^0 q_c + p^1 C_{cd} \, p^d \right) 
\\ 
&& 
-\frac{ (S + \bar S) }{\pi} \left( 12 \left(Y^0 - {\bar Y}^0 \right)^2  + \left( \ln \eta^{24} (S) \right)' 
(S + \bar S) \left({\bar Y}^0 \right)^2 
+ \left( \ln \eta^{24} ({\bar S}) \right)' (S + \bar S) \left(Y^0 \right)^2 \right) \;, \nonumber
\end{eqnarray}
where in this expression the derivatives are with respect to $S$ and to $\bar S$, respectively.

Next, let us relate \eqref{res-q01-sum} to the free energy of a macroscopic black hole.  To this end, we first note
that the mixed ensemble \eqref{osv-pf} involves summing \eqref{res-q01-sum} over $q_a$.
The macroscopic free energy, which corresponds to a critical point of the free energy functional, is obtained by extremizing the exponent in \eqref{res-q01-sum} with respect to $q_a$.  Performing this extremization we find
\begin{equation}\label{backeq}
\tilde{\phi}^a = -\frac{2 C^{ab}q^B_b}{S+\bar{S}}\,,
\end{equation}
where $\tilde{\phi}^a = \phi^a + i p^a (S - \bar S)/(S + \bar S)$.  Then, inserting \eqref{backeq} into the exponent of \eqref{res-q01-sum} gives
\begin{eqnarray}
{\cal F}_E (p, \phi) = 
\frac{1}{4} (S + \bar S) \left[ p^a C_{ab} \, p^b - \phi^a C_{ab} \phi^b - 2 i \frac{S - \bar S}{S + \bar S}
\, \phi^a C_{ab} p^b \right] + 4  \, \Omega(S, \bar S) \;.
\label{bh-free-energy}
\end{eqnarray}
When $l^0=l^1=0$, 
the value $q_a^B$ can be thought of as a background
charge that defines an attractor background geometry in view of the fact that \eqref{backeq}
is simply the attractor equation for the real part of the scalar moduli fields $Y^a$, cf. \eqref{eq:phia-attrac}.
Then, \eqref{bh-free-energy} equals the macroscopic free energy of this background charge black hole \cite{LopesCardoso:2006bg}
\begin{eqnarray}
{\cal F}_E (p, \phi) = 4 \left[ \Im F^{(0)}(Y) + \Omega(Y, \bar Y) \right]\Big|_{Y^I = \tfrac12 (\phi^I + i p^I)}  \;,
\end{eqnarray} 
and the sum over the $q_a$ can be interpreted as a sum over fluctuations about this 
attractor background.
In these expressions, $Y^0, \bar Y^0, S$ and $\bar S $ are defined with shifts $\phi^0$ and $\phi^1$,
as in \eqref{eq:SbSl}. When $l^0=l^1=0$, $S$ and $Y^0$ become related to the attractor values for a single-centered black hole. 
Indeed, using \eqref{bh-free-energy}, we 
can rewrite \eqref{res-q01-sum} as \begin{eqnarray}
\label{resultq0q1}
\sum_{q_0, q_1} d(q,p) \, {\rm e}^{\pi q_I \phi^I   } 
&=& 
\sum_{l^0 \in \mathbb{Z}, l^1 \in \mathbb{Z}} \; 
\frac{M}{(S + \bar S)^2 \, \left(Y^0 {\bar Y}^0 \right)^2} \; 
\nonumber\\
&& 
\qquad  {\rm exp} \left[ \pi \, {\cal F}_E (p, \phi) 
+ \frac{\pi}{S + \bar S} \, V^a \, C_{ab} \, V^b \right] ,
\end{eqnarray}
where
\begin{equation}
V^a = C^{ab} q_b + \tfrac12(\phi^a (S + \bar S) + i p^a (S - \bar S) ) \;.
\label{eq:v-comb}
\end{equation}
where $V^a$ describes a fluctuation about the background charge \eqref{backeq}.  This follows by writing $V^a$ as 
\begin{equation}\label{v1-comb}
V^a= C^{ab}(q_b - q^B_b)= C^{ab}\delta q_b \;,
\end{equation}
where we used \eqref{backeq}.
One can see that the fluctuations can be space-like, time-like or null due to the hyperbolic structure of the charge-lattice metric $C_{ab}$. In fact, if one thinks of the exponent as a free energy functional used to define the action for a  partition function in a discrete hyperbolic lattice, then it is easy to see that there are no extrema of the action, but only critical points corresponding to  single-centered black holes, since at any given point on this hyperbolic lattice there is always a space-like and a time-like direction. 

For the purpose of single-centered black hole entropy, we will only be interested in fluctuations in the $l^0=l^1=0$ sector. 
The appearance of the other sectors in the mixed partition function function can be explained by analyzing the microcanonical degeneracy given in 
\begin{equation}
d(Q,P,R)=\int \int \int \dd \sigma \, \dd \rho \, \dd v \, \frac{{\rm e}^{\pi i (Q \sigma + P \rho + R (2v-1) )}}{\Phi_{10}(\rho,\sigma,v)} \;.
\label{triple-int}
\end{equation}
Here the contours are chosen such that the imaginary parts of the three arguments are fixed at certain values determined in terms of the charge invariants
\cite{Cheng:2007ch}, and the real parts are chosen to run from $0$ to $1$.
The integrand has second order poles corresponding to ${\cal D}_{(n,m,p)}$.  In order to evaluate
the residues at these poles, one can use the invariance of the integrand under imaginary translations in 
$\sigma, \rho$ and $v$ to map ${\cal D}_{(n,m,p)}$ to ${\cal D}_{(0,0,0)}$ while extending the range
of the real parts of $\sigma, \rho$ and $v$ to the real line.
In the case of $\Phi_{10} (\sigma(\theta), \rho(\theta), \theta)$, ${\cal D}_{(n,m,p)}$ are mapped
to ${\cal D}_{(n,m,0)}$. This involves an extension of the range of the real part of $\theta$ and a 
simultaneous translation in $\phi^0$ and $\phi^1$ in order to preserve the ranges of 
$\sigma(\theta)$ and $\rho(\theta)$, cf. \eqref{eq:srvthet}.  Here, the values of $(l^0, l^1)$
mod $(p^1, p^0)$
correspond to the increase in the ranges of the real parts of $\sigma$ and $\rho$ in $\Phi_{10} (\sigma, \rho, v)$.
The integral over $v$ is done by expressing the leading divisor as a function of $v$ and then evaluating the residue. The remaining two integrals are then performed by saddle point integration. The contour that passes through the saddle point is chosen in such a way that the two variables become conjugate to each other along the contour and that at the saddle point they correspond to the heterotic axion-dilaton pair and its conjugate
\cite{LopesCardoso:2004xf}. Another way of expressing this is to say that the axion and dilaton scalars become real on this specific contour.
 
The triple integral \eqref{triple-int} also helps in defining background charges $q_0$ and $q_1$, as follows. 
The imaginary parts of the integration variables, for the single-centered degeneracy, are expressed in terms of T-duality invariants as \cite{Cheng:2007ch}
\begin{eqnarray}
\Im \sigma &= & -2 \Lambda \frac{P}{| {\mbf Q}\wedge {\mbf P}|}\,,\nonumber\\
\Im \rho &=& - 2 \Lambda \frac{Q}{ |{\mbf Q}\wedge {\mbf P}|}\,,\nonumber\\
\Im v &=& 2 \Lambda \frac{R}{|{\mbf Q}\wedge {\mbf P}|}\,.
\end{eqnarray}
Using the definitions of the period matrix variables \eqref{eq:srvthet}
in terms of $\phi^0$ and $\phi^1$, and the definition of $q_0$ and $q_1$ in terms of $Q$ and $P$ (cf. \eqref{QPqq}),
respectively, we get an expression  relating the background values of $q_0$ and $q_1$ to 
$\phi^0$, $\phi^1$ and the other background charges,
determined up to a positive constant $\Lambda$.

\subsection{Attractor geometry constraints on $q_a$ summation}

Summarizing,  by summing over $(q_0, q_1)$, the number of integrations in \eqref{triple-int}
gets reduced from three to one, and the remaining integral over $\theta$ can be evaluated via
residues.  This is achieved by introducing an infinite sum over integers $l^{0,1}\in \mathbb{Z}$, which makes the shift symmetry
$\phi^{0,1} \rightarrow \phi^{0,1} + 2 i $ manifest.  In the above we assumed that $p^0 p^1 \neq 0$.  The result \eqref{res-q01-sum}
remains valid when setting either $p^0=0$ or $p^1=0$.  This can be checked (and we will do so in the next subsection) by
redoing the above calculations using instead $(Q, R)$ and $(P, R)$ as summation variables, following \cite{Shih:2005he}.

Eq. \eqref{res-q01-sum}  captures part of the OSV partition function for single-center black holes, 
namely the part associated with $n_2 =1$. This yields the leading contribution to the partition function.
Next, we would like to sum over charges $q_a$.  Here one faces the problem that
one has to restrict to states with 
${\rm sgn}(\varrho_a C^{ab}\varrho_b) > 0$,.
Implementing this condition in a 
sum over charges $q_a$ is somewhat unwieldy. Note that this constraint becomes trivial in the rigid limit. Namely, when 
decoupling gravity, we recover a low-energy gauge theory based on a prepotential $F^{(0)}$
with a definite metric $C_{ab}$, and the associated sum over the electric charges is unrestricted.
To proceed, we note that 
 a simplification occurs when setting $p^0 = 0$, since in this case
$\varrho_a C^{ab} \varrho_b = - (p^1)^2 \, P$  and 
${\rm sgn}(\varrho_a C^{ab} \varrho_b) = {\rm sgn} (-P)$, which only depends on magnetic charges.  
Further, to make contact with a gravity partition function over single-centered black holes, one notes that black holes with $p^0=0$ have a near horizon geometry that, at an appropriate point of the moduli space, can be  seen as a BTZ excitation of $AdS_3$ \cite{Nampuri:2007gw}. The OSV ensemble naturally sums over the $q_I$ charges and keeps the $p^I$ charges  fixed. The fixed charges precisely define the $AdS_3$ background while the summed charges define excitations in this background.  For these reasons, we will restrict ourselves to a summation over states with  $p^0 =0$ in the remainder of this paper.

\subsection{Case $p^0=0$: summing over $q_a$}

We will now compute the OSV mixed partition function \eqref{osv-pf}  for the case when $p^0 =0$.

First, we redo the steps leading to \eqref{res-q01-sum} for the case $p^0 =0$.  Using
\begin{eqnarray}
Q &=& 2 q_0 p^1 - q_a C^{ab} q_b \;, \nonumber\\
R &=& - p^1 q_1 + p^a q_a \;, \nonumber\\
P &=& - p^a C_{ab} p^b \;,
\label{def:QPR}
\end{eqnarray}
we convert the sum over $(q_0, q_1)$ into a sum over $(Q, R)$ and obtain \cite{Shih:2005he}
\begin{eqnarray}
\sum_{q_0, q_1} d(q,p) \, {\rm e}^{\pi q_I \phi^I   } 
&=&\frac{1}{ (p^1)^2} \sum_{\tiny \begin{array}{c} l^0, l^1= 0 , \dots |p^1| -1
\end{array}}\; 
\, \sum_{Q, R} \,  d(Q, P, R) \, \nonumber\\
&&{\rm exp}\left[ \frac{\pi \hat{\phi}^0}{2 p^1} \left(Q + q_a C^{ab} q_b \right) -
\frac{\pi \hat{\phi}^1}{p^1} \left(R  - p^a q_a \right) 
+ \pi q_a \phi^a 
\right] \;, 
\label{eq:QPsump0}
\end{eqnarray}
where now $P$ is independent from $Q$ and $R$.  We set
\begin{eqnarray}
\sigma_* &=& \frac{i \hat{\phi}^0}{2 p^1} = \frac{i}{S + \bar S} \;, \nonumber\\
v_* &=& \frac12 - \frac{i \hat{\phi}^1}{2 p^1} = \frac{S}{S + \bar S} \;,
\end{eqnarray}
where we recall from \eqref{eq:SbSl}, 
\begin{eqnarray}
Y^0 &=& {\bar Y}^0 = \frac12 \hat{\phi}^0 = \frac12 \left( \phi^0 + 2 i l^0 \right)\;, \nonumber\\
S &=& \frac{- i \hat{\phi^1}+ p^1}{\hat{\phi}^0} = 
 \frac{- i \phi^1 + p^1 + 2 l^1 }{\phi^0 + 2 i  l^0} \;, \nonumber\\
\bar S &=& \frac{i \hat{\phi}^1 + p^1 }{\hat{\phi}^0} = 
\frac{i \phi^1 + p^1 - 2 l^1}{\phi^0 + 2i  l^0} \;.
\label{Y0S-p0}
\end{eqnarray}
Once again, observe that $\bar{Y}^0$ and $\bar S$ are not the complex conjugates of $Y^0$
and $S$ when $l^{0,1}$ are non-vanishing.
Next, we use 
the definition
\begin{eqnarray}
\sum_{Q, R} \,  d(Q, P, R) \, {\rm e}^{- 2 \pi i \left( \tfrac12 Q  \sigma_* + R ( v_* - \tfrac12) \right) }
= \int_0^1 \dd \rho_1 \, \frac{{\rm e}^{i \pi  P \rho}}{\Phi_{10} (\sigma_*, \rho, v_*)} \;,
\label{def-sumQR}
\end{eqnarray}
where $\rho = \rho_1 + i \rho_2$, and $\rho_2$ is fixed.  We obtain \cite{Shih:2005he}
\begin{eqnarray}
\sum_{q_0, q_1} d(q,p) \, {\rm e}^{\pi q_I \phi^I   } 
&=&\frac{1}{ (p^1)^2} \sum_{\tiny \begin{array}{c} l^0, l^1= 0 , \dots |p^1| -1
\end{array}}\; 
\,\int_0^1 \dd \rho_1 \, \frac{1}{\Phi_{10} (\sigma_*, \rho, v_*)} 
 \nonumber\\
&&{\rm exp}\left[ - i \pi \, \sigma_* \, q_a C^{ab} q_b  +
\pi q_a \left( \phi^a + \frac{\hat{\phi}^1}{p^1} \, p^a \right) 
+ i \pi  P \rho
\right] \;.
\label{eq:interm}
\end{eqnarray}
We consider single-center black hole solutions, so that $P < 0$.
As before, identifying $d(q,p)$ with the microcanonical dyonic degeneracy generated by $\frac{1}{\Phi_{10}}$ fixes
the arguments $(\sigma_*, \rho, v_*)$ to satisfy the Siegel upper-half plane conditions 
\begin{eqnarray}
{\rm Im} \, \sigma_* &=& \frac{\phi^0}{2 p^1} > 0 \;, \nonumber\\
{\rm Im} \, \rho &>& 0 \;, \nonumber\\
{\rm Im} \, \sigma_* \, {\rm Im} \, \rho &>& ( {\rm Im} \, v_* )^2 = \left(\frac{\phi^1}{2 p^1}\right)^2 \;.
\label{Siegel-cond}
\end{eqnarray}
Next, we proceed as in the previous subsection.  
Using the characterization of the zeroes of
$\Phi_{10}$ corresponding to $n_2 =1$ in terms of integers $n,m,p$ we extend the sum
over $l^{0,1}$ to run over all the integers, and we extend the range of integration to
$- \infty < \rho_1 < \infty$.  Using ${\cal D} = \sigma_* (\rho - \rho_*)$ with $\rho_* = 
(v^2_* - v_*)/\sigma_* = i S \bar S/(S + 
\bar S)$
as well as 
$\Phi_{10} \approx {\cal D}^2 \, \Delta$, 
we obtain the analogue of \eqref{eq:M-expl}, 
\begin{eqnarray}
M = - (S + \bar S)^2 (Y^0)^2  \left[ P + \pi^{-1} \left( 
    \ln \eta^{24} (S) \right)' 
+ \pi^{-1} \left( \ln \eta^{24} ({\bar S}) \right)' \right] \;.
\label{eq:def-check-M}
\end{eqnarray}
where we used $p^1 = Y^0 ( S + \bar S)$.  Eventually, we obtain for the unregularized OSV partition function (up to an overall numerical constant),
\begin{eqnarray}
\label{osvpf0}
\cZ_{\rm OSV} (p, \phi) &=& 
\sum_{l^0 \in \mathbb{Z}, l^1 \in \mathbb{Z}}  \;
\frac{\left[ P + \pi^{-1} \left( 
    \ln \eta^{24} (S) \right)' 
+ \pi^{-1} \left( \ln \eta^{24} ({\bar S}) \right)' \right] }{(Y^0)^2}
\,
{\rm e}^{ 2 \pi i \, \tau_m \, \sfQ_m (p) 
+ 4 \pi \Omega(S, \bar S) } \nonumber\\
&& \qquad \qquad \qquad \qquad  \sum_{q_a} e^{2 \pi i \, \tau_e \, \sfQ_e (q) + 2 \pi i \, \sfB_e (z,q)} \;,
\end{eqnarray}
where 
\begin{eqnarray}
\tau_m &=& i \frac{S {\bar S}}{S + \bar S}   \;\;\;,\;\;\;
\tau_e = \frac{i}{S + \bar S} \;, \nonumber\\
A_{ab} &=& - C_{ab} \;\;\;,\;\;\; A^{ab} = - C^{ab} \;, \nonumber\\
\sfQ_m (p) &=& \frac12 \, p^a A_{ab} p^b \;\;\;,\;\;\;
\sfQ_e (q) = \frac12 \, q_a A^{ab} q_b \;\;\;,\;\;\;
\sfB_e(z, q) = z_a A^{ab} q_b \;, \nonumber\\
z_a &=& \frac{i}{2} C_{ab} \left( \phi^b + 
\frac{i (S - \bar S)}{S + \bar S} \, p^b \right)
= a_a \, 
\tau_e \, + b_a \;, 
\label{valuez-p0}
\end{eqnarray}
where 
$a = {\rm Im} \, z/ {\rm Im } \, \tau$ and $b = {\rm Im} \, (\bar z \, \tau)/{\rm Im} \, \tau$.
Here $\sfQ_m (p)$ and $\sfQ_e (q)$ are indefinite quadratic forms, and $\sfB_e (z,q)$
is the bilinear form associated to $\sfQ_e (q)$.  Using $\tau_e = \sigma_*$ we obtain 
that $\tau_e$ takes values in the complex
upper half plane by virtue of the Siegel upper half plane conditions \eqref{Siegel-cond}.
Note that  \eqref{osvpf0} agrees with \eqref{res-q01-sum} when setting $p^0 =0$.

For generic values of $l^0$ and $l^1$
both
$a_a$ and $b_a$ are non-vanishing in the decomposition \eqref{valuez-p0}.
On the other hand, when $l^0=l^1=0$,
$(S + \bar S)$ and $i(S - \bar S)$ are both real, and hence $b_a=0$.  We will return to this issue
in the next subsection when regularizing the sum \eqref{osvpf0}.

The matrix $A_{ab}$ has signature $(n-2,1)$, and hence the quadratic form $\sfQ_e(q)$ is indefinite, rendering
the sum \eqref{osvpf0} over $q_a$ divergent, as discussed previously.    
We propose to regulate the divergence by 
turning the sum 
over the $q_a$ in 
\eqref{osvpf0}
into an indefinite theta function $\vartheta(z; \tau_e)$ following \cite{Zwegers:2008}.

\subsection{Regularizing $\cZ_{\rm OSV}(p, \phi)$} \label{OSVpoisson}

Recall that in our particular model, $A^{ab}$ is integer valued and indefinite, 
and $\tau_e$ takes values on the upper half complex plane by (\ref{Siegel-cond}). This is precisely the setting where indefinite theta functions can be defined. We will now modify the definition of the OSV sum in order to obtain an indefinite theta function, and discuss the consequences of this procedure. A physically motivated discussion of the regulatory procedure is given in Appendix \ref{toy model} via a toy model. The main properties of indefinite theta functions are summarized in the Appendix \ref{app-ind-theta-fun}. An indefinite theta function \eqref{ind-theta} differs from an ordinary theta function by the presence of an extra factor $\rho$ which deals with the indefinite directions, preserving modular properties. This factor $\rho$ explicitly depends on two vectors $c_1$ and $c_2$. Depending on the specific form of $\rho$ these two vectors are used to project out the lattice points giving an exponentially growing contribution, or to weight them with a positive definite quadratic form. 

Thus, by introducing the weight $\rho_{(e)}$ in the OSV partition function (\ref{osvpf0}), we obtain a convergent and modular sum
\begin{eqnarray}
 \label{osvpf0-reg}
\cZ_{\rm OSV}^{c_1 , c_2} (p, \phi) &=& 
\sum_{l^0 \in \mathbb{Z}, l^1 \in \mathbb{Z}}  \;
\frac{\left[ P + \pi^{-1} \left( 
    \ln \eta^{24} (S) \right)' 
+ \pi^{-1} \left( \ln \eta^{24} ({\bar S}) \right)' \right] }{(Y^0)^2}
\, \nonumber\\
&& \qquad \qquad \qquad \qquad {\rm e}^{ 2 \pi i \, \tau_m \, \sfQ_m (p) 
+ 4 \pi \Omega(S, \bar S) } 
\vartheta(z; \tau_e) \;. 
\end{eqnarray}
Note that this should be intended as part of the \textit{definition} of the electric sum, as we are not going to remove the weight $\rho_{(e)}$ in the following. Whether this factor can be derived from first principle, and not just by macroscopic arguments, is clearly an interesting question. 

Having obtained a modular object\footnote{There is a subtlety in the sector $l^0=l^1=0$, to which we will return 
at the end of this subsection.}, we now consider the modular transformation $(\tau_e, z) \rightarrow (- 1/\tau_e, z/\tau_e)$.
Using that $A_{ab}$ is integer valued,
$\vartheta(z; \tau_e)$ transforms as \cite{Zwegers:2008}
\begin{eqnarray}
\vartheta (z/\tau_e ; - 1/\tau_e) &=& \frac{1}{\sqrt{- \det A}} \, 
\left(-i \tau_e\right)^{(n-1)/2} \, {\rm e}^{2 \pi i \, \sfQ_e (z)/\tau_e } \, \vartheta(z; \tau_e) 
\nonumber\\
&=& 
\sum_{\nu \in \mathbb{Z}^{n-1}} \rho_{(e)} (\nu + {\tilde a} ; - 1/\tau_e) \, {\rm e}^{- 2 \pi i \,\sfQ_e (\nu)/\tau_e + 2 \pi i \,\sfB_e (z/\tau_e, \nu)} 
\;,
\end{eqnarray}
where
\begin{equation}
\tilde{a} = \frac{{\rm Im} (z/\tau_e)}{{\rm Im}(-1/\tau_e) } = \frac{{\rm Im} (b/\tau_e)}{{\rm Im}(-1/\tau_e) } = 
- b\;.
\end{equation}
Hence we obtain 
\begin{eqnarray}
\vartheta (z; \tau_e) &= &\frac{\sqrt{- \det A}}{(-i\tau_e)^{(n-1)/2}} \,  {\rm e}^{- 2 \pi i \, \sfQ_e(z)/\tau_e } 
\sum_{\nu_a \in \mathbb{Z}^{n-1}} \rho_{(e)} (\nu - b  ; - 1/\tau_e) \, {\rm e}^{- 2 \pi i \, \sfQ_e (\nu)/\tau_e + 2 \pi i \, \sfB_e (z/\tau_e, \nu)} 
\nonumber\\
&=&
\frac{\sqrt{- \det A}}{(-i \tau_e)^{(n-1)/2}} \,  
\sum_{\nu_a \in \mathbb{Z}^{n-1}} \rho_{(e)} (\nu + b  ; - 1/\tau_e) \, 
 {\rm e}^{- 2 \pi i \, \sfQ_e (z + \nu )/\tau_e } \;.
\label{theta-el-mod}
\end{eqnarray}
Observe that 
\begin{eqnarray}
z_a + \nu_a &=& \frac{i}{2} C_{ab} \left( \hat{\phi}^b + i \frac{(S - \bar S)}{S + \bar S} \, p^b
\right)
\;, \nonumber\\
\hat{\phi}^a &=& \phi^a - 2 i C^{ab} \, \nu_b \;,
\label{hat-phi-sh}
\end{eqnarray}
which makes it manifest that \eqref{osvpf0-reg} has 
the shift symmetry $\phi^a \rightarrow \phi^a + 2 i$.

Using \eqref{theta-el-mod}, $\cZ^{c_1 , c_2}_{\rm OSV} (p, \phi)$ gets expressed as
\begin{eqnarray}
\label{reg-osvpf-free}
\cZ^{c_1 , c_2}_{\rm OSV} (p, \phi) &=& 
\sqrt{- \det A} 
\sum_{l^0 \in \mathbb{Z}, l^1 \in \mathbb{Z}}  \;
(S + \bar S)^{(n-1)/2} \,
\frac{\left[ P + \pi^{-1} \left( 
    \ln \eta^{24} (S) \right)' 
+ \pi^{-1} \left( \ln \eta^{24} ({\bar S}) \right)' \right] }{(Y^0)^2}
\nonumber\\
&& \sum_{\nu \in \mathbb{Z}^{n-1}} \, {\rm e}^{\pi \, {\cal F}_E (p, \hat{\phi})} \rho_{(e)} (\nu + b  ; - 1/\tau_e) 
\;,
\end{eqnarray}
where ${\cal F}_E (p, \hat{\phi})$ denotes the free energy \eqref{bh-free-energy} with $\phi^a$ replaced by
$\hat{\phi}^a$. 

Apart from modular transformations, the indefinite theta function may also be subjected
to elliptic transformations.  One such transformation is induced by the S-duality
transformation $S \rightarrow S + i \, \lambda$ with $\lambda \in \mathbb{Z}$.
This transformation induces the shift
$z_a \rightarrow z_a + \lambda_a \tau_e$, where $\lambda_a = - C_{ab} \, p^b \lambda$,
as can be seen from \eqref{valuez-p0}.  Under this transformation, 
the indefinite theta function picks up a factor \cite{Zwegers:2008}
${\rm exp} [ - 2 \pi i \, \tau \, {\sfQ}(\lambda_a) - 2 \pi i {\sfB}(z, \lambda_a)]$.
This particular elliptic transformation can also be viewed as inducing a shift of  the background
charge $q^B_a$ given in \eqref{backeq}.  Namely, using \eqref{valuez-p0}, the 
above transformation can also be obtained by performing the shift  $\phi^a  \rightarrow 
\phi^a - 2 \lambda \, p^a/(S + \bar S)$ which, using \eqref{backeq}, translates into  shifting 
the background charge by
$q^B_a \rightarrow q^B_a + \lambda C_{ab}p^b$. 
This shows how the background charge dependence is encoded in the elliptic transformation.

Introducing $T^a$ as  
\begin{eqnarray}
\left(T + \bar T \right)^a = \frac{(S + \bar S)}{p^1} \, p^a \;,
\label{TpbT}
\end{eqnarray}
and using $p^1 = Y^0 (S + \bar S)$ we get
\begin{equation}
P = - \left(T + \bar T \right)^a C_{ab}  \left(T + \bar T \right)^b \, (Y^0)^2 
\end{equation}
as well as 
\begin{eqnarray}
\cZ^{c_1 , c_2}_{\rm OSV} (p, \phi) &=& 2 \, 
\sqrt{- \det A} \, 
\sum_{l^0 \in \mathbb{Z}, l^1 \in \mathbb{Z}}  \; 
\frac{ (S + \bar S)^{(n-3)/2}}{(Y^0)^2 } \; \nonumber\\
&& \left[\hat{K} + 4 (S + \bar S)^2 \partial_S \partial_{\bar S} \Omega \right]
 \; \sum_{\nu \in \mathbb{Z}^{n-1}} \, {\rm e}^{\pi \, {\cal F}_E (p, \hat{\phi})} \rho_{(e)} (\nu + b  ; - 1/\tau_e) 
\;,
\label{reg-osv-p0}
\end{eqnarray}
where (we recall that here $Y^0 = \bar Y^0$)
\begin{equation}
\hat{K} = \frac12 Y^0 \bar{Y}^0 (S + \bar S) \left[ 
\left(T + {\bar{T}}\right)^a 
C_{ab} \left({T} + {\bar{T}}\right)^b + 4 \frac{\partial_S \Omega}{(Y^0)^2}
+ 4 \frac{\partial_{\bar S} \Omega}{(\bar{Y}^0)^2}
\right] \;.
\end{equation}
This quantity equals the K\"ahler potential $K = i \left({\bar Y}^I \, F_I - Y^I \, {\bar F}_I \right)$ computed
from $F = F^{(0)} + 2 i \Omega$ (where $F^{(0)}$ and $\Omega$ are given in \eqref{eq:heterotic-prep} and in 
\eqref{rel-theta-st}), with  $T^a = - i Y^a/Y^0$ replaced by \eqref{TpbT}, and
with $Y^0$ and $S$ replaced by the shifted quantities \eqref{Y0S-p0}.
Observe that $K$ is invariant under both S- and T-duality \cite{LopesCardoso:2006bg}.  This extends to 
$\hat{K}$ and to the combination
$\hat{K} + 4 (S + \bar S)^2 \partial_S \partial_{\bar S} \Omega$, provided S- and T-duality are defined in
the same way when acting on the shifted fields \eqref{Y0S-p0} and \eqref{TpbT}.
Note also that if we artificially take $n=27$
(which corresponds to taking a model with 28 abelian gauge fields, just as in the original $\cN=4$ 
model), the term $(S + \bar S)^{12}$ in \eqref{reg-osv-p0}
precisely cancels against a similar term
coming from $\Omega(S, \bar S)$ in \eqref{rel-theta-st} \cite{deWit:2007dn}, so that \eqref{reg-osv-p0} becomes
\begin{eqnarray}
\cZ^{c_1 , c_2}_{\rm OSV} (p, \phi) &=& 2 \, 
\sqrt{- \det A} 
\sum_{l^0 \in \mathbb{Z}, l^1 \in \mathbb{Z}}  \; 
\left[\hat{K} + 4 (S + \bar S)^2 \partial_S \partial_{\bar S} \Omega \right] \\ && \times \nonumber
 \sum_{\nu \in \mathbb{Z}^{n-1}} \,  {\rm e}^{ F_{\rm holo} (\hat{\phi}^a , p^a , S ) - \ln Y^0}  {\rm e}^{ \bar{F}_{\rm holo} (\hat{\phi}^a , p^a , \bar{S} ) - \ln Y^0} \, \rho_{(e)} (\nu + b  ; - 1/\tau_e) \ ,
 \label{Z27}
\end{eqnarray}
where
\begin{eqnarray}
F_{\rm holo} (\hat{\phi}^a , p^a , S ) &=& - \frac{\pi}{4} \, S \, \left(\hat{\phi}^a + i p^a \right) C_{ab}  \left(\hat{\phi}^b+ i p^b \right) 
- \ln \eta^{24}(S) \;, \nonumber\\
{\bar F}_{\rm holo} (\hat{\phi}^a , p^a , \bar{S} )  &=& - \frac{\pi}{4} \, {\bar S} \, \left(\hat{\phi}^a - i p^a \right) C_{ab}  \left(\hat{\phi}^b- i p^b \right) 
- \ln \eta^{24}(\bar S) \;.
\end{eqnarray}
Thus, \eqref{Z27} takes a form reminiscent of $|{\rm e}^{F_{\rm top}}|^2$ (where $F_{\rm top}$
denotes the holomorphic topological free energy), 
with an additional duality invariant measure factor \cite{LopesCardoso:2006bg} as well an extra weight factor $\rho_{(e)}$.

Our result for the regularized OSV partition function \eqref{reg-osv-p0}
contains a sum over indefinite theta functions over different $(l^0, l^1)$ sectors. 
In each $(l^0, l^1)$-sector, we can choose wedge vectors $c_1$ and $c_2$ to define the regulating error functions.
Let us consider the sector $l^0=l^1 =0$ in more detail, and let us discuss a subtlety to which we already alluded to
above. The sector $l^0=l^1 =0$ describes the 
semi-classical sector, and hence we must demand that our choice of wedge vectors and regularization yields sensible results in the semi-classical regime. The exact semi-classical point corresponds to $\nu = 0$ in the $l^0 = l^1 = 0$ sector. However, as already mentioned, we have $b_a=0$ in this sector, and hence, both error functions in the regulator vanish. To resolve this conundrum, we propose a shift in the fluctuation \eqref{v1-comb}
\begin{equation}
V^a\rightarrow V^a +\lim_{\lambda\rightarrow 0} i \lambda U^a\,,
\end{equation}
 and, through \eqref{eq:v-comb}, this automatically yields a shift in $\phi^a$ as 
\begin{equation}
\phi^a \rightarrow \phi^a + \lim_{\lambda\rightarrow 0} 2 i \lambda U^a/(S+\bar{S}).
\label{phi-U-sh}
\end{equation}
This modifies the definition of $b_a$ at the semi-classical point to 
\begin{equation}
b_a = -\frac{\lambda C_{ab}U^b}{(S+\bar{S})} \;.
\end{equation}
Here, we have used the fact that at the semi-classical point,  $\Re(S-\bar{S}) = \Im(S+\bar{S}) = 0$. The modified free energy has an extra term $\frac{-\lambda^2 {\sfQ}(U)}{S+\bar{S}} -  i \lambda U^a C_{ab} \tilde{\phi}^b$, as can be seen
from \eqref{bh-free-energy}.
We will now pick appropriate value of $c_1$ and $c_2$ to preserve the classical free energy.

In a well defined classical limit one should ensure that the exponential corrections coming from the error function do not affect the free energy. From (\ref{reg-osv-p0}) both the exponential and $\rho$ go as $S+\bar{S}$. If we denote $y = -\im \frac{1}{\tau_e}$ and $x_1$ and $x_2$ the remaining factors in the error function, we can write 
\begin{equation}
\rho = E (x_1 \, \sqrt{y}) - E (x_2 \sqrt{y}) \;.
\end{equation}
Here, since we are interested in the semi-classical limit, we set $l^0 = l^1 = 0$ and obtain
\begin{eqnarray}
x_1 &=& \frac{ c^1_a A^{ab} (\nu_b+b_b)}{\sqrt{- {\sf Q}(c_1)}} \;, \nonumber\\
x_2 &=& \frac{ c^2_a A^{ab} (\nu_b+b_b)}{\sqrt{- {\sf Q}(c_2)}} \;,
\end{eqnarray}
which are invariant under rescaling of the $c_i$. In the semi-classical limit we have 
$(S+\bar{S})\rightarrow \infty$, and hence we consider the expansion of the error function 
 \eqref{rho-x2} around $x=0$ as \cite{Gradshteyn},
\begin{equation}
 E (x) \simeq 2 \, {\rm e}^{-\pi x^2} \sum_{n=0}^{\infty}  \, \frac{(2 \pi)^n \, 
 x^{2n+1}}{(2n+1)!!} = 2 \, {\rm e}^{-\pi 
 x^2} \, x + \dots.
\end{equation}
In our case this yields (setting $\nu_a =0$)
\begin{equation}\label{exprho}
\rho \simeq 2 \, {\rm e}^{\pi \, 
\frac{\lambda^2 (c^1_a U^a)^2}{{\sf Q}(c_1) (S+\bar{S})}}\frac{\lambda \, c^1_a U^a}{\sqrt{- {\sf Q}(c_1) \,  
(S+\bar{S})}}
\left(1- \frac{c^2_a U^a \sqrt{-{\sf Q}(c_1)}}{c^1_b U^b \sqrt{- {\sf Q}(c_2)}} \, {\rm e}^{\pi 
\big[
\frac{\lambda^2 (c^2_a U^a)^2}{{\sf Q}(c_2) (S+\bar{S})}-\frac{\lambda^2 (c^1_a U^a)^2}{{\sf Q}(c_1) (S+\bar{S})}\big]} 
\right)+ O(\lambda^2) \;.
\end{equation}
We  choose $U^a$ to be a spacelike vector with norm $U = \sqrt{U^a C_{ab} U^b}$, so that in some T-duality frame we can bring it to the form
$U^a = (1,\frac12 U^2,\vec{0})$, where vector $\vec{0}$ spans the timelike $SO(r-2)$ directions and the non-zero slots fill out a hyperbolic lattice. Then, in this frame, we choose $c^1_a = (\frac12 U^2,1,\vec{0})$ and $c^2_a = (1,\frac12 U^2, \vec{0})$, so as to ensure that
the exponent in the regulator outside the brackets in \eqref{exprho} fully cancels the extra real term in the free energy $\pi {\cal F}_E$
. In addition, for large $U$, the second term in the brackets is subleading compared to the first (when 
$\lambda \rightarrow 0$). Then, normalizing the regulator in \eqref{exprho} by $\lambda^{1 + \alpha} \,{\rm e}^{- \pi i \lambda U^a C_{ab} \tilde{\phi}^b}$, where $0 < \alpha < 1$,
and demanding that $\lim_{\lambda\rightarrow 0,S+\bar{S}\rightarrow \infty}\lambda^{2 \alpha} (S+\bar{S})$ stays finite, we remove all  leading order contributions to the semi-classical free energy.

Let us now briefly comment on the symmetries of the regularized OSV partition function.
The indefinite theta function has modular and elliptic transformation properties.  The modular transformation
$(\tau_e, z) \rightarrow (-1/\tau_e, z/\tau_e)$ implements Poisson resummation, which we employed to extract the
semi-classical free energy, see \eqref{reg-osv-p0}.  The elliptic transformation
$z_a \rightarrow z_a + \lambda_a \tau_e$, with $\lambda_a = - C_{ab} \, p^b \lambda$ and $\lambda \in \mathbb{Z}$,
induces a shift of the background charge, 
$q^B_a \rightarrow q^B_a + \lambda C_{ab}p^b$. This is a reflection of the underlying S-duality invariance 
that is present in the original $N=4$ theory, see \eqref{eq:electro-magn-dual_charges}. The regularized OSV partition function also has the shift
symmetry $\phi^I \rightarrow \phi^I + 2i$, which expresses 
integrality of the charges. This shift symmetry was made manifest using the following steps.
In going from the OSV partition function 
\eqref{osv-pf}
to the dyonic degeneracy formula \eqref{eq:deg-sum}, the summation variables 
changed from charges to T-duality invariant charge bilinears.  The chemical potentials appearing in the 
dyonic degeneracy formula \eqref{eq:deg-sum}
have translational symmetries associated with 
the integrality of the invariant
charge bilinears. 
In order to ensure the stronger condition for the integrality of charges, we had to make the translation symmetry of the original OSV potentials explicit by introducing new dummy summation variables and complexifying the potentials. Hence, the shift symmetry $\phi^I \rightarrow \phi^I + 2i$ of \eqref{osv-pf} is manifest in the result
\eqref{reg-osv-p0}, which is entirely expressed in terms of hatted potential $\hat{\phi}^I$ that are complex.  This was achieved by 
regularizing the sum over $q_a$ by turning it into an indefinite theta function $\vartheta(z; \tau)$, and subsequently
applying a modular transformation to it.  To achieve this, we had to introduce a vector $U^a$ with norm ${\sf Q}(U^a) < 0$.
We may identify $U^a$ with the real part of an asymptotic $T^a$-modulus lying in the K\"ahler cone.
This shows that, in general, only an $SO(1, r)$ subgroup of the T-duality symmetry group
is preserved
by the regulator/the specific choice of the vectors $c_1$ and $c_2$ that enter in the definition of
$\rho$, in the Zweger's prescription.

\subsection{$\cZ^{c_1 , c_2}_{\rm OSV} (p, \phi)$  from an expansion of $\frac{1}{\Phi_{10}}$ in powers of $P$ }

In the OSV ensemble, the $p^I$ are kept fixed, which implies that 
 when $p^0=0$, the charge bilinear $P$ is  constant.
   This allows us to obtain an exact expression for $\cZ^{c_1 , c_2}_{\rm OSV} (p, \phi)$
 by using a Fourier expansion of $1/\Phi_{10}$ in $P$-modes, as follows.

We consider the expansion \cite{Dabholkar:2012nd}
\begin{equation}
\frac{1}{\Phi_{10} (\sigma , \rho , v)} = \sum_{m \geq -1} \psi_m (\sigma, v) \, {\rm e}^{2 \pi i m \rho} \;,
\end{equation}
which converges for ${\rm Im} \, \rho >0$.  Inserting it 
into \eqref{def-sumQR} selects the coefficient $m = - P/2$ and yields,
\begin{eqnarray}
\label{res-q01-sum-p0}
\sum_{q_0, q_1} d(q,p) \, {\rm e}^{\pi q_I \phi^I   } 
&=& \frac{1}{(p^1)^2} 
\, \sum_{l^0, l^1 = 0, \dots, |p^1|-1} \psi_{-P/2} (\sigma_*, v_*) \\
&& \hskip - 25mm
{\rm exp} \left[
\frac{\pi}{S + \bar S} \left( q_a C^{ab} q_b + (S + \bar S) q_a \phi^a + 
i (S - \bar S) q_a p^a  \right) 
\right] \;.
\nonumber
\end{eqnarray}
Observe that $\psi_{-P/2} (\sigma_*, v_*)$ is invariant under shifts $\phi^{0,1} \rightarrow
\phi^{0,1} + 2 i p^1 n$ with $n \in \mathbb{Z}$, since $\sigma_* \rightarrow \sigma_* -n$,
$v_* \rightarrow v_* + n$, which constitutes an invariance of $\Phi_{10}$.  It follows that
\eqref{res-q01-sum-p0} is invariant under shifts $l^{0,1} \rightarrow l^{0,1} + 1$.

Proceeding as above, we regularize the sum over the $q_a$ by turning it into an indefinite theta function, 
\begin{eqnarray}
\cZ_{\rm OSV}^{c_1 , c_2} (p, \phi) = 
\frac{1}{(Y^0 (S + \bar S))^2} 
\, \sum_{l^0, l^1 = 0, \dots, |p^1|-1} \psi_{-P/2} (\sigma_*, v_*) \, \vartheta(z; \tau_e) \;,
\label{psitheta}
\end{eqnarray}
with $z_a$ given as in \eqref{valuez-p0}.
Performing the modular transformation \eqref{theta-el-mod} we get
\begin{eqnarray}
\cZ_{\rm OSV}^{c_1 , c_2} (p, \phi) &=& \sqrt{- \det A} \, 
\frac{(S + \bar S)^{(n-5)/2}}{(Y^0)^2} 
\, \sum_{l^0, l^1 = 0, \dots, |p^1|-1} \psi_{-P/2} (\sigma_*, v_*) \nonumber\\
&& \sum_{\nu_a \in \mathbb{Z}^{n-1}} \rho_{(e)} (\nu + b  ; - 1/\tau_e) \, 
 {\rm e}^{- 2 \pi i \, \sfQ_e (z + \nu )/\tau_e } \;.
 \label{exp-free}
\end{eqnarray}

\section{Conclusions}

In this paper we have streamlined a new approach to deal with black hole partition functions in quantum gravity.  The main feature is that the indefinite character of the charge lattice, a distinctive feature of gravity, and the need to preserve as many symmetries as possible, point towards the necessity of new mathematical structures to deal with the sums over microscopic states. We propose that the theory of indefinite theta functions may play a distinctive role in this program, as elucidated below.

\subsection{Partition functions and divergences}

In order to contextualize the divergences in the mixed ensemble, it is instructive to analyze the counting formula in two other ensembles. 
The first ensemble is defined by chemical potentials corresponding to the variation of the T-duality charge bilinears, where the partition function is given by 
\begin{equation}
\frac{1}{\Phi_{10}(\sigma , \rho , v)}= \sum_{Q\leq 2,P\leq 2,\, R\,\in\,\mathbb{Z}} d(Q,P,R) 
\, {\rm e}^{- \pi i (Q \, \sigma+ P \, \rho + R \, (2 \, v-1) )} \; .
\end{equation}
In the Siegel upper-half plane, where this series is well-defined, $\Im \sigma \gg 1\,,\Im \rho >\gg 1$ and $\Im \rho \Im \sigma \gg (\Im v)^2$, and hence the $Q$ and $P$ expansions are convergent. However, for a given value of $\Im v$, the series is divergent in the $R$-sum, as the $R$ sum goes over both positive and negative values. Hence, one cannot define the series for both positive and negative values of $R$, for a fixed value of $\Im v$. This is related to the meromorphic structure of $\Phi_{10}$, arising from its double zero structure, which we dealt with, in detail, in section \ref{sec2}. In particular, the partition function has a double pole at $\frac{y}{(1-y)^2}$, where $y= {\rm e}^{- 2\pi i v}$. One can expand this series about $y=0$ or $y = \infty$, resulting in an expansion in positive powers of $y$ , corresponding to positive R or an expansion in negative powers of $y$, corresponding to negative powers of $R$, respectively. But one cannot analytically continue from one expansion to another due to the pole at $y=1$ corresponding to $\Im v=0$. As one moves through the pole, one picks up the residue around the pole, and this results in a jump in the degeneracy across a line of marginal stability corresponding to the appearance or disappearance of decadent dyons. One way to extract the microcanonical degeneracy  and regulate the series is to compute the degeneracy for one sign of $R$, corresponding to the fixed sign of $\Im v$, which makes the series well-defined. Then, we define the degeneracy of the charge configuration with the same value of $Q$ and $P$, but with the opposite sign of $R$, as being equal to the degeneracy 
of the computed charge configuration, by applying parity-invariance \cite{dabholkar:2007vk}.

A second ensemble in which we can write down a dyonic counting formula is obtained by fixing $P$ and varying the other two charge bilinears.
We can write down a partition function in this ensemble by going to a point in the Siegel upper-half plane where $\Im \rho \gg \Im \sigma$. This allows us to expand the Siegel form as \cite{Dabholkar:2012nd}
\begin{equation}
\frac{1}{\Phi_{10}(\sigma , \rho , v)}= \sum_{m \geq -1} \psi_{10,m} ( \sigma, v) \, {\rm e}^{ 2 \pi i m \rho} \;,
\end{equation}
where $\psi_{10,m}$ is a Jacobi form of weight $10$ and index $m$. 
The Jacobi form inherits its meromorphicity from the Siegel modular form, and this leads to a divergence at the double poles. One can regulate this divergence by splitting the Jacobi form into two mock modular forms, one of which, the polar part, $\psi^P$, encodes the double poles of the Jacobi form, and hence the jumps across the lines of marginal stability due to the decadent dyons, and the other is the finite analytic part of the Jacobi form which counts the immortal single-centered dyons, $\psi^F$, as 
\begin{equation}
\psi_{10,m}= \psi^P + \psi^F \;.
\end{equation}

On the other hand, in the  mixed statistical ensemble, the partition function is written as 
$\cZ_{\rm OSV} (p^I, \phi^I) = \sum_{q_I}d(q,p) \, {\rm e}^{\pi q_I \phi^I}$.
For any sign of the chemical potentials, as we sum over both positive and negative values of $q_I$, this series is divergent. As we saw in section \ref{sec2}, this divergence  shows up in the evaluation of the fluctuations about the critical point contribution to the free energy of the partition function. 
In this paper we have physically motivated a reason to use a soft regulatory mechanism to handle the divergences in the partition function written in the OSV ensemble (see Appendix \ref{toy model}). We used a soft regulator following Zwegers to convert this into an indefinite theta function. The regulator converts the divergent series into an indefinite theta function. However, this regulator could be one of many choices to define a convergent series. In order to provide a physical basis for it, we notice that the resulting well-defined partition function counts single-centered black holes. Hence, a physical justification for our regulator lies in establishing a  connection between the regulated partition function in the mixed ensemble, and the finite mock modular form. Accordingly, one must turn to the connection between the theory of indefinite theta functions and mock modular forms analyzed in \cite{Zwegers:2008},  and show that the mock modular form associated with this indefinite theta function is precisely the one that encodes the single-centered degeneracies.   
This connection can be worked backwards in theories like the STU or FHSV models, where there is a strong 
  suggestion \cite{Manschot:2009ia,Manschot:2010xp}
that the wall crossing phenomena are encoded in terms of indefinite theta functions, to be able to extract mock modular forms and hence, partition functions for counting single-center black holes. These open questions are being currently pursued.

\subsection{Summary and context}

The simple idea described and implemented in this paper provides a starting point to define OSV-like sums over single-centered black hole microstates which have a semiclassical limit consistent with supergravity. Indeed, the need of having sums over states which are both mathematically meaningful (and not just formal) and compatible with an infrared gravitational description has been the guiding principle of our approach. 
One can extend the procedure implemented in this paper to the  $n_2>1$ poles of the Siegel modular form that counts the microscopic states in the theory.
Regularizing the OSV partition function
using indefinite theta functions renders it convergent
while maintaining the shift symmetry 
\eqref{eq:shift-q-p}, and allows to make contact with semi-classical results through the usage
of modular transformations that implement Poisson resummation.

In order to prove the absolute convergence of the indefinite theta function series, one can show 
\cite{Zwegers:2008}
that the series of the absolute values of the terms in the theta function converges faster than a canonical series obtained by effectively modifying the metric ${\sfQ}(c)$ of the charge lattice so as to remove the indefinite direction, and hence regulate the series. The norms of the vectors then change to ${\sf Q}(\nu) \rightarrow {\sf Q}(\nu) - \frac{{\sf B}(c,\nu)^2}{2 {\sf Q}(c)}$ (cf. \eqref{Qc}). Hence one could alternatively choose to simply modify the metric as above. It turns out that this is equivalent to introducing a canonical partition function regulator ${\rm e}^{-\beta H}$, where $H$ is the Hamiltonian of the dyonic system seen as a bound state of D-branes in the theory.  
Indeed, if we consider the exponent $Q_c(\nu) \, {\rm Im} \tau_e$ in \eqref{Qc}  and focus on the second term, given by
$B(c,\nu)^2/[- 2 (S + \bar S) Q(c)]$ (here we set $l^0=l^1=0$), and perform the replacements $\beta = 1/(S + \bar S)$ and 
$u^a = C^{ab} c_b/\sqrt{-2 Q(c)}$, we obtain $\beta \, (\nu_a u^a)^2$ with $C_{ab} u^a u^b =1$.  This is precisely
the H-regulator proposed in (6.15) of \cite{Dabholkar:2005dt}. Thus, amusingly, the H-regulator used in \cite{Dabholkar:2005dt,Denef:2007vg} can be formally identified with the second term of $Q_c$. We note, however, that the H-regulator
has, so far, only been used at strong topological
string coupling, 
c.f. (2.51) and (2.54) in \cite{Denef:2007vg}. We are not aware of its extension to weak topological string coupling,
which corresponds to the semi-classical limit.

One can also view the present work in the context of the picture of the quantum entropy function introduced by Sen in 
\cite{Sen:2008vm}.  The quantum entropy function, which counts the microstates of a supersymmetric black hole, was defined in the $AdS_2$ background which formed the near horizon geometry of the black hole. In this background, the fluctuation over the various fields had to be performed keeping the charge fixed since, in two dimensions, the charge is associated with the non-normalizable part of the electric field. Hence, the quantum gravity partition function computed in this background is bound to be the microcanonical partition function of black hole microstates, and it can be expressed as the exponential of the Legendre transform of the full quantum action evaluated on this background with respect to the charges of the black hole, to give the full quantum entropy function.
On the other hand, to compute a canonical partition function, we looked at black holes which, at some point in their moduli space, have a near horizon background factor of $AdS_3$. The central charge of the holographically dual CFT and the radius of $AdS_3$ are fixed by the $p^I,$ while the chiral excitation that defines the black hole is determined by the $q_I$\footnote{Strictly speaking, the chiral excitation is proportional to the spectral flow invariant constructed from the $q_I$.}. 
Hence, the partition function is defined by summing over the $q_I$ while keeping $p^I$ fixed. This mixed ensemble counts fluctuations in $AdS_3$, and the free energy computed in this ensemble will include not just the single-center black hole excitations, but also other excitations of the $AdS_3$ vacuum.  The associated partition function can be asymptotically expressed as the exponent of a free energy which is the Legendre transform of the action with respect to the $q_I$. 

Finally, the work presented here paves the way for rigorously defining black hole partition functions in a grand canonical ensemble.
The grand canonical ensemble sums over every charge in the system \cite{LopesCardoso:2006bg} and can be thought of as summing over all fluctuations. The free energy computed in this case will have contributions from various $AdS_3$ 
backgrounds, each of which defines a mixed ensemble. This quantity will be the full 
Euclidean quantum action that encode the dynamics
of black hole backgrounds in the theory. 
A well defined properly regulated partition function in the grand canonical ensemble is therefore quintessentially important in an understanding of
the underlying stringy effective action of the theory. 

Finally, we note that our results indicate the need for defining indefinite theta functions on more general lattices.

\vskip 5mm

\subsection*{Acknowledgements}
We acknowledge valuable discussions with
Atish Dabholkar, Jan de Boer, 
Bernard de Wit, Jan Manschot, 
Thomas Mohaupt, Sameer Murthy, Nicolas Orantin,  
Alvaro Osorio, Sara Pasquetti, Jan Troost and Marcel Vonk.
The work of G.L.C. and M.C. is
partially supported by the Center for Mathematical Analysis, Geometry
and Dynamical Systems (IST/Portugal), a unit of the LARSyS laboratory, as well as 
by Funda\c{c}\~{a}o para a Ci\^{e}ncia e a Tecnologia
(FCT/Portugal) through grants CERN/FP/116386/2010 and PTDC/MAT/119689/2010 and EXCL/MAT-GEO/0222/2012. 
R. J. gratefully acknowledges the support of the Gulbenkian Foundation through the scholarship
program
{\it Programa Talentos em Matem\'atica 2011-12.}
M. C. is also supported by the by Funda\c{c}\~{a}o para a Ci\^{e}ncia e a Tecnologia
(FCT/Portugal) via the Ci\^{e}ncia2008 program. This work is also partially supported by the COST action
MP1210 {\it The String Theory Universe}.

\appendix

\section{S-duality, attractor equations, and the Hesse potential \label{sdual-attrac}}

In this Appendix we will review some of the duality properties of the effective $\cN=2$ description and their relation with the attractor equations. In particular we will also discuss the relation between the effective free energy and the Hesse potential. Consider the prepotential \eqref{eq:heterotic-prep}.
Associated to each $Y^I$ is a pair $(q_I, p^I)$ of electric/magnetic charges.  This is the 
charge vector in the so-called type IIA polarization.  It is related to the one in the heterotic polarization by
\begin{align}
\tilde q = &\, (q_0, - p^1, q_a) \;, \nonumber\\
\tilde p = &\, (p^0, q_1, p^a ) \;.
\label{qpvec}
\end{align}
Under S-duality, 
\begin{equation}
        \begin{array}{rcl}
      Y^0 &\to& d \, Y^0 + c\, Y^1 \;,\\
      Y^1 &\to& a \, Y^1 + b \, Y^0 \;,\\
      Y^a &\to& d\, Y^a - c \,C^{ab} \, F_b  \;,
    \end{array}
    \quad
    \begin{array}{rcl}
      F_0 &\to& a\,  F_0 -b\,F_1 \;, \\
      F_1 &\to& d \,F_1 -c\, F_0 \;,\\
      F_a &\to& a \, F_a -b\,  C_{ab} \, Y^b \;,
    \end{array}
    \label{eq:electro-magn-dual}
\end{equation}
where $a, b, c, d$ are real parameters that satisfy $a d - b c = 1$. It acts as 
\begin{equation}
S \rightarrow \frac{ a S - i b }{i c S + d} 
\end{equation}
on $S = - i Y^1/Y^0$, and as follows on the charges, 
\begin{equation}
        \begin{array}{rcl}
      p^0 &\to& d \, p^0 + c\, p^1 \;,\\
      p^1 &\to& a \, p^1 + b \, p^0 \;,\\
      p^a &\to& d\, p^a - c \,C^{ab} \, q_b  \;,
    \end{array}
    \quad
    \begin{array}{rcl}
      q_0 &\to& a\,  q_0 -b\,q_1 \;, \\
      q_1 &\to& d \,q_1 -c\, q_0 \;,\\
      q_a &\to& a \, q_a -b\,  C_{ab} \, p^b \;.
    \end{array}
    \label{eq:electro-magn-dual_charges}
\end{equation}
In particular, under the transformation $a=d=0, b = - c = 1$, we have $S \rightarrow 1/S$ and 
$q_a \rightarrow - C_{ab} p^b \;,\; p^a \rightarrow C^{ab} q_b$.

We define 
\begin{eqnarray}
\phi^I &=& Y^I + {\bar Y}^I \;, \nonumber\\
\chi_I &=& F^{(0)}_I + {\bar F}^{(0)}_I \;,
\end{eqnarray}
where $F_I^{(0)} = \partial F^{(0)} / \partial Y^I$ and ${\bar F}^{(0)}_I = \partial {\bar F}^{(0)} / \partial {\bar Y}^I$.  Observe that the combinations $q_0 \phi^0 + q_1 \phi^1$ and 
 $p^0 \chi_0 + p^1 \chi_1$ are invariant under S-duality transformations
\eqref{eq:electro-magn-dual} and \eqref{eq:electro-magn-dual_charges}.

The attractor equations relating the $Y^I$ to the charges $(q_I, p^I)$ are 
\begin{eqnarray}
Y^I - {\bar Y}^I &=& i \, p^I\;, \nonumber\\
F^{(0)}_I - \bar{F}^{(0)}_I & = & i \, q_{I} \;,
\label{eq:attrac-eqs}
\end{eqnarray}
where $F_I^{(0)} = \partial F^{(0)} / \partial Y^I$ and ${\bar F}^{(0)}_I = \partial {\bar F}^{(0)} / \partial {\bar Y}^I$.

Consider the prepotential \eqref{eq:heterotic-prep}.
Imposing the magnetic attractor equations,
\begin{equation}
Y^I = \tfrac12 \left[
\phi^I  + i p^I \right] \;,
\label{Y-magn-att}
\end{equation}
as well as the electric attractor equations for the $q_a$, leads to a full determination of the $Y^I$ in 
terms of $S$, 
\begin{equation}
S = - i \frac{Y^1}{Y^0} = \frac{- i \phi^1 + p^1}{\phi^0 + i p^0} \;,
\label{S-phi-p}
\end{equation}
as follows \cite{LopesCardoso:2006bg},
\begin{equation}
Y^0 = \frac{\bar{P}(\bar S)}{S + \bar S} \;\;\;,\;\;\; 
Y^1 = i \, \frac{S \, \bar{P}(\bar S)}{S + \bar S} \;\;\;,\;\;\;
Y^a = - \frac{C^{ab} {\bar Q}_b (\bar S) }{ S + \bar S} \;,
\label{eq:back-val-Y}
\end{equation}
where 
\begin{eqnarray}
P(S) & =& p^1 - i S p^0 \;, \nonumber\\
Q_b (S) &=& q_{b} +  i \, S\,C_{bc} \, p^c \;.
\label{Q-P}
\end{eqnarray}
Using \eqref{eq:back-val-Y},
the attractor values for $(\phi^a, \chi_a)$ are
\begin{eqnarray}
&&\phi^a + 2\, \frac{C^{ab} \, q_b}{S + {\bar S}} +  \frac{i(S - \bar S)}{S + \bar S} \, p^a= 0 \;, 
\label{eq:phia-attrac}
\end{eqnarray}
and
\begin{equation}
\chi_a -2 \frac{|S|^2}{S + \bar S}  C_{ab} p^b
-
i \frac{(S - \bar S)}{S + \bar S} q_a  = 0 \;.
\label{eq:chia-attrac}
\end{equation}
Using  \eqref{eq:heterotic-prep} and \eqref{Q-P} we compute
\begin{eqnarray}
\chi_0 &=& \frac{i}{2 (S + \bar S)} \left[ \frac{S \, {\bar Q}_a C^{ab} {\bar Q}_b}{\bar{P}(\bar S)}
- 
\frac{\bar S \, {Q}_a C^{ab} {Q}_b}{P(S)} 
\right] \;, \nonumber\\
\chi_1 &=& - \frac{1}{2 (S + \bar S)} \left[ \frac{{\bar Q}_a C^{ab} {\bar Q}_b}{\bar{P}(\bar S)}
+
\frac{ {Q}_a C^{ab} {Q}_b}{P(S)} 
\right] \;.
\label{values-chi01}
\end{eqnarray}

Next, let us consider the macroscopic free energy based on \eqref{eq:heterotic-prep}.  It is given by
\eqref{bh-free-energy} with $\Omega =0$ (and $l^0 = l^1 = 0$),
\begin{eqnarray}
 {\cal F}^{(0)}_E (p, \phi) &=& 4  \left[ \Im F^{(0)}(Y) \right]\Big|_{Y^I = \tfrac12 (\phi^I + i p^I)} 
\nonumber\\
&=&
\tfrac14 (S + \bar S) \left[ p^a C_{ab} \, p^b - \phi^a C_{ab} \phi^b - 2 i \frac{S - \bar S}{S + \bar S}
\, \phi^a C_{ab} p^b \right]  \;.
\label{free-ener-F0}
\end{eqnarray}
The modular parameters $\tau_e$ and $\tau_m$
that appear in \eqref{valuez-p0}
can be defined starting 
from the macroscopic free energy \eqref{free-ener-F0}, as follows.
We use the attractor equation \eqref{eq:phia-attrac} to express 
$\phi^a$ in terms of $S, \bar S$ and the charges $(p^a, q_a)$.
Then, viewing ${\cal F}^{(0)}_E$ as a function of 
 $p^a, q_a$ (and $S, \bar S$) we get
\begin{eqnarray}
\frac{\partial^2 {\cal F}^{(0)}_E}{\partial q_a \partial q_b} = - \frac{2}{S + \bar S} \, C^{ab} \;\;\;,\;\;\;
\frac{\partial^2 {\cal F}^{(0)}_E}{\partial q_a \partial  p^b} = 0 \;\;\;,\;\;\; 
\frac{\partial^2 {\cal F}^{(0)}_E}{\partial  p^a \partial  p^b} = \frac{2 |S|^2}{S + \bar S} \, C_{ab} \;.
\end{eqnarray}
If we now set 
\begin{eqnarray}
\frac{\partial^2 {\cal F}^{(0)}_E}{\partial q_a \partial q_b} = 2 i \tau_e\,  C^{ab} \;\;\;,\;\;\;
\frac{\partial^2 {\cal F}^{(0)}_E}{\partial  p^a \partial  p^b} = - 2 i \tau_m\, C_{ab} \;,
\end{eqnarray}
we obtain $\tau_e = i /(S + \bar S)$ and $\tau_m = i |S|^2/(S + \bar S)$.

Finally, let us consider the Hesse potential $H^{(0)}(\phi, \chi, S, \bar S)$ that is obtained by Legendre transformation of
the free energy \eqref{free-ener-F0} with respect to $p^a$,
\begin{equation}
H^{(0)}(\phi, \chi, S, \bar S) = {\cal F}^{(0)}_E (p, \phi) - p^a \, \chi_a \;.
\end{equation}
Using $\chi_a = \partial {\cal F}^{(0)}_E / \partial p^a$ we obtain
\begin{equation}
H^{(0)}(\phi, \chi, S, \bar S) = - \frac{1}{S + \bar S} \left[ \chi_a C^{ab} \chi_b + |S|^2 \phi^a C_{ab} \phi^b 
+ i (S - \bar S) \phi^a \chi_a \right] \;.
\label{hesse-0}
\end{equation}


\section{Toy model for a regulator \label{toy model}}

We outline a simple model of regulating a divergent series so as to be able to provide a physical motive for the  prescription of Zwegers.
Consider the series $\sum_{V} {\rm e}^{V^2}$, where $V_a$ is a vector in a two-dimensional Lorentzian space,
and $V^2 = V_a C^{ab} V_b$. The spacelike vectors render this series divergent. One simple way to regulate this series in a Lorentz-invariant fashion is simply to sum over only the time-like vectors. To implement this we go to the light cone gauge, where the norm-squared of the vector $V_a$ is given by  $V^2 = V_+ V_-$, and for time-like vectors we have ${\rm sgn}(V_+)=- {\rm sgn}(V_-)$, whereas
${\rm sgn}(V_+)= {\rm sgn}(V_-)$ for space-like vectors. Hence inserting the regulator ${\rm sgn}(V_+)-{\rm sgn}(V_-)$, we see that the series is regulated in a Lorentz-invariant fashion.
This hard-regulator method depends on only counting the vectors in regions 1 and 3 wedged between the null-vectors. As a soft regulatory mechanism, one could choose a wedge of non-null vectors $c^1$ and $c^2$. Then following the argument of the sign-function, we see that the argument of the regulatory function is $\frac{V_a C^{ab} c_a}{\sqrt{c_a C^{ab} c_b}}$. In the null limit, where $c_a C^{ab} c_b$ approaches null, this function should revert to the ${\rm sgn}$ function. Using the limiting identity for the error function, $\lim_{k\rightarrow\infty} E(k x) = {\rm sgn}(x)$, we see that in general the regulator could be chosen to be $E(\frac{V_a C^{ab} c^1_{a}}{\sqrt{c^1_a C^{ab} c^1_b}})-E(\frac{V_a C^{ab} c^2_a}{\sqrt{c^2_a C^{ab} c^2_b}})$. This is precisely the regulatory proposal of Zwegers to define indefinite theta functions, which we review below.  Note that in the case of a regulator based on error functions, all the vectors $V_a$ contribute to the sum,
but those that would cause a divergence now appear with an exponential damping
factor, rendering the sum convergent.  Thus, no vectors are discarded in this case.

\section{Properties of indefinite theta functions\label{app-ind-theta-fun}}

In this Appendix we review various properties of indefinite theta functions.
Indefinite theta functions $\vartheta (z;\tau)$
were defined and studied by Zwegers in \cite{Zwegers:2008},
and are modified versions of the sums considered in \cite{GottZag1996}.
They have found recent string theory applications
in \cite{Manschot:2010xp,Alim:2010cf,Alexandrov:2012au,Manschot:2012rx,Dabholkar:2012nd}.  
Indefinite theta functions are based on quadratic forms 
$\sfQ:
\mathbb{R}^r \rightarrow \mathbb{R}$ of signature  $(r-1,1)$, 
defined in terms of symmetric non-degenerate $r \times r$ matrices $A$ with integer coefficients, 
$\sfQ(x) = \frac{1}{2}x^T A x$.  The associated bilinear form $\sfB$ is 
$\sfB(x,y)=x^T A y = \sfQ(x+y) - \sfQ(x) - \sfQ(y)$.
The convergence is implemented by the presence of an additional factor $\rho$  in the sum
defining $\vartheta (z; \tau)$,
\begin{equation}
\vartheta (z; \tau) = \sum_{n \in \mathbb{Z}^r} \rho (n + a; \tau) \, {\rm e}^{ 2\pi i \,\tau \, \sfQ(n) + 2 \pi i \, \sfB(n,z)} \;,
\label{ind-theta-fun}
\end{equation}
where $\tau \in {\cal H}$ takes values in the complex upper half plane ${\cal H}$, and $z \in \mathbb{C}^r$, with $a, b \in \mathbb{R}^r$ defined by $z = a \, \tau + b$. The factor $\rho$ is the difference of two functions $\rho^c$, 
$\rho = \rho^{c_1} - \rho^{c_2}$. The $\rho^{c_i}$ depend on real vectors $c_i \in \mathbb{R}^r$ 
that satisfy $\sfQ(c_i) \leq 0$.  In the following, we take $\sfQ(c_i) <0$ for both $c_1$ and $c_2$. The other possibility, that is when $\sfQ (c_i) = 0$, will be discussed in Appendix \ref{thetaexample}. The set of vectors with $\sfQ(c) <0$ has two components, and we take $c_1$ and $c_2$ to be in the same component, so 
that $\sfB(c_1, c_2)<0$. Then, the
$\rho^{c_i}$ are given in terms of error functions,
\begin{equation}
\rho^c (n + a; \tau) = E \left(\frac{\sfB(c, n + a )}{\sqrt{- \sfQ(c)}} \, \sqrt{{\rm Im} \, \tau} \right) \;,
\label{rho-x}
\end{equation}
where 
\begin{equation} \label{rho-x2}
E(x) = 2 \int_0^x {\rm e}^{- \pi u^2}  \dd u  = {\rm sgn}(x)\;\left( 1 - \beta(x^2) \right) \;\;\;,\;\;\; x \in \mathbb{R} \;,
\end{equation}
with
\begin{equation}
\beta(x^2) = \int_{x^2}^{\infty} u^{-1/2} \, {\rm e}^{- \pi u } \, \dd u \;.
\end{equation} 
Observe that $\rho^c (n + a; \tau) $ is non-holomorphic in $\tau$.
Also note that the definition of $\rho(n + a; \tau)$ doesn't change if we replace $c_i$ by $\lambda \, c_i$, with 
$\lambda \in \mathbb{R}_+$.  This implies that two $c_i$ belonging to the same component of $\sfQ(c) <0$ should not
be collinear, since otherwise $\rho =0$, and that we may replace the condition $\sfQ(c_i) <0$ by $\sfQ(c_i) = -1$
\cite{Zwegers:2008}.

As shown in \cite{Zwegers:2008}, \eqref{ind-theta-fun} is convergent and has nice modular 
and elliptic transformation properties that are similar to those of theta functions based
on positive definite quadratic forms.  In the following, we briefly highlight various aspects
that go into proving these remarkable facts.  Following  \cite{Zwegers:2008}, we consider
the indefinite theta function $\vartheta_{a,b} (\tau) $ defined by
\begin{equation}
\vartheta_{a,b} (\tau) = {\rm e}^{2 \pi i \sfQ(a) \tau + 2 \pi i \sfB(a,b) } \, \vartheta(z; \tau) 
=
\sum_{\nu \in a + \mathbb{Z}^r} \, \rho(\nu; \tau) \,  \e^{2 \pi i \sfQ(\nu) \tau + 
2 \pi i \sfB(\nu, b)} \;.
\label{theta-series}
\end{equation}
We begin by sketching the proof of convergence of \eqref{theta-series}.
Since the proof is rather lengthy,
we focus on 
the regulator $\tilde{\rho}(x) = - {\rm sgn}(x) \, \beta(x^2)$, which is related to $\rho$ according to
\eqref{rho-x2}.
To prove the convergence of the series using  $\tilde{\rho}$, we will need the following two lemmata
from \cite{Zwegers:2008}.
The first lemma states that
\begin{equation}
0 \le \beta(x^2) \le \e^{- \pi x^2} \;,
\end{equation}
$\forall \, x \in \mathbb{R}$.
To show this, we consider the function $f(x) = \beta (x^2) - {\rm e}^{- \pi x^2}$.  
It vanishes at $x =0, \pm \infty$.  Away from these values, it has 
extrema located at $\pi x = {\rm sgn}(x) $, which are local minima.  Hence
it follows that $f(x) \leq 0$, which establishes the lemma.

The second lemma that is needed states that the combination
\begin{equation}
\sfQ_c(\nu) := \sfQ(\nu) - \frac{\sfB(c,\nu)^2}{2 \sfQ(c)}
\label{Qc}
\end{equation}
is positive definite. Here, $c$ is a vector satisfying $\sfQ(c) <0$.  Note that $\sfQ_c(\nu) >0$ regardless
of the sign of $ \sfQ(\nu)$.
This lemma can be proven as follows.
First let us consider the case when 
$\nu \in \mathbb{R}^r$ is linearly independent of $c$.  Then 
the quadratic form $\sfQ$ has signature $(1,1)$ on the two-span$\{c,\nu\}$, and hence the matrix
\begin{eqnarray}
 \left( \begin{array}{ c c }
  2\sfQ(c)  & \sfB(c,\nu) \\
   \sfB(c,\nu) & 2\sfQ(\nu)  \\
\end{array} \right)
\end{eqnarray} 
has determinant $< 0$, so noting that $\sfQ(c)<0$ we obtain
\begin{equation}
4\sfQ(c)\sfQ(\nu) - \sfB^2(c,\nu) < 0 \leftrightarrow \sfQ_c(\nu)  > 0 \;.
\end{equation}
On the other hand, when $\nu = \lambda \, c$ with $\lambda \neq 0$, we obtain $\sfQ_c(\nu) = - \sfQ(c) \, \lambda^2 > 0$,
which shows that the combination \eqref{Qc} is always positive definite.

Next, 
using $\sfQ(c) < 0$, we compute
\begin{eqnarray}
\big|\tilde{\rho}(\nu; \tau) \,  \e^{2 \pi i \sfQ(\nu) \tau + 
2 \pi i \sfB(\nu, b)} \big| 
\le \e^{\pi \frac{\sfB^2(c,\nu)}{\sfQ(c)}\, {\rm Im} \, \tau} \big| \e^{2 \pi i \sfQ(\nu) \tau + 2 \pi i \sfB(\nu, b)}
\big| = \e^{-2 \pi \sfQ_c(\nu) \, {\rm Im} \, \tau } \;.
\label{series-Qc}
\end{eqnarray}
Since $\sfQ_c (\nu) >0$, the series 
\begin{equation}
\sum_{\nu \in a +  \mathbb{Z}^r} \e^{-2 \pi \sfQ_c(\nu) \, {\rm Im} \, \tau } 
\end{equation}
 converges, and thus $\vartheta_{a,b}(\tau)$ is absolutely convergent for the above choice of
 regulator, and hence convergent. It is also uniformly convergent for ${\rm Im} \tau  \geq \varepsilon > 0$.
 The proof of convergence for the regulator $\rho$ 
 is much more involved, but proceeds along similar lines \cite{Zwegers:2008}.

Now let us consider the behavior of $\vartheta_{a,b}(\tau)$ under the modular transformation
$\tau \rightarrow - 1/\tau$.  The proof given in \cite{Zwegers:2008}
establishing that $\vartheta_{a,b}(\tau)$ has a good behavior under this transformation 
requires the regulator $\rho(\nu; \tau)$
to be
an odd function of $\nu$, and the derivatives $\partial \rho/\partial \nu$ to exist.
In \cite{Zwegers:2008} it is shown that under
$\tau \rightarrow - 1/\tau$, $\vartheta_{a,b}(\tau)$ transforms as 
\begin{equation}
\vartheta_{a,b}(-1/\tau) = \frac{i}{\sqrt{-\det A}}(-i \tau)^{r/2} \, {\rm e}^{2 \pi i  \sfB(a,b)} 
\, \sum_{p \in A^{-1} \mathbb{Z}^r \; {\rm mod} \;\mathbb{Z}^r} 
\vartheta_{b + p, -a} (\tau) \;.
\label{transf-theta-tau}
\end{equation}
We now sketch the proof leading to this result.  It uses a lemma as well as Poisson resummation.
The lemma states that for all $\alpha \in \mathbb{R}^r$ and $\tau \in {\cal H}$, 
\begin{equation}
\int_{\mathbb{R}^r} \rho(a; \tau) \, \e^{2 \pi i \sfQ(a) \tau + 2 \pi i \sfB(a,\alpha)} \dd a 
= \frac{1}{\sqrt{- \det A}} \frac{i}{(-i \tau)^{r/2}} \, \rho(\alpha; - 1/\tau) \,
 \e^{- 2 \pi i \sfQ(\alpha)/\tau} \;.
 \label{lemma-theta}
\end{equation}
Its proof goes as follows.  The integral on the left hand side is convergent. 
Using
\begin{equation}
\frac{\partial}{\partial \alpha_l} \e^{2 \pi i \sfQ(a \tau + \alpha)/\tau}=\frac{1}{\tau} \frac{\partial}{\partial a_l} \e^{2 \pi i \sfQ(a \tau + \alpha)/\tau} \;,
\end{equation}
we obtain
\begin{equation}
\frac{\partial}{\partial \alpha_l} \left( \e^{2 \pi i \sfQ(\alpha) /\tau }
\int_{\mathbb{R}^r} \rho(a; \tau) \, \e^{2 \pi i \sfQ(a) \tau + 2 \pi i \sfB(a,\alpha)} \dd a \right)
= - \frac{1}{\tau} \int_{\mathbb{R}^r} 
\frac{\partial \rho}{\partial a_l} (a; \tau) \, \e^{2 \pi i \sfQ(a \tau + \alpha) /\tau } \dd a \;,
\label{dif-rho-int}
\end{equation}
where we integrated by parts and used that the boundary terms do not contribute.
Since $\rho$ is the difference of two error functions, 
the derivatives $\frac{\partial \rho}{\partial a_l} (a; \tau)$ yield derivatives of error functions.
We therefore consider the following expression which appears on the right hand side of the above equation,
\begin{equation}
\int_{\mathbb{R}^r}
E' \left(\frac{\sfB(c,a)}{\sqrt{-\sfQ(c)}} \sqrt{{\rm Im} \, \tau} \right) \, \e^{2 \pi i \sfQ(a \tau + \alpha) /\tau } \dd a 
= 2 \e^{2 \pi i \sfQ(\alpha) /\tau }
\int_{\mathbb{R}^r} {\rm e}^{\pi \frac{\sfB^2(c, a)}{\sfQ(c)} {\rm Im} \, \tau} \, {\rm e}^{ 2 \pi i \, \sfQ(a) \tau + 2 \pi i \sfB(a, \alpha)} \dd a
\;,
\label{res-inter-Epr}
\end{equation}
where we used the property $E'(x) = 2 {\rm e}^{- \pi x^2}$.
As shown in  \cite{Zwegers:2008}, the integral over $\mathbb{R}^r$ can be split into an integral over $\mathbb{R}$
(corresponding to the direction associated with the negative eigenvalue of $\sfQ$) and an integral over
$\mathbb{R}^{r-1}$ associated with the directions that correspond to the positive eigenvalues of $\sfQ$.
The integration over  $\mathbb{R}$ would yield a divergent result where it not for the presence
of the additional factor $E'$ which converts the factor $\tau$ appearing in the
exponent on the right hand side into a factor $\bar \tau$, rendering the integration
over  $\mathbb{R}$ well behaved, 
\begin{equation}
\int_{\mathbb{R}} \e^{2 \pi i \sfQ(c) \, a^2_c \, \bar \tau  + 4 \pi i \sfQ(c) a_c \alpha_c} \dd a_c \;,
\end{equation}
where we refer to \cite{Zwegers:2008} for a detailed derivation of
this remarkable result.

Using this, the right hand side of \eqref{res-inter-Epr} evaluates to
\begin{eqnarray}
&& \sqrt{{\rm Im} \, \tau} \, 
\int_{\mathbb{R}^r}
E' \left(\frac{\sfB(c,a)}{\sqrt{-\sfQ(c)}} \sqrt{{\rm Im} \, \tau} \right) \, \e^{2 \pi i \sfQ(a \tau + \alpha) /\tau } \dd a 
\nonumber\\
&& \qquad = \frac{\sqrt{{\rm Im} (-1/\tau)}}{(-i \tau)^{r/2-1}} \frac{1}{\sqrt{- \det A}} \, 
E' \left(\frac{\sfB(c,\alpha)}{\sqrt{-\sfQ(c)}} {\sqrt{{\rm Im} (-1/\tau)}} \right) \;.
\end{eqnarray}
Using this expression in \eqref{dif-rho-int} leads to
\begin{equation}
\frac{\partial}{\partial \alpha_l} \left( \e^{2 \pi i \sfQ(\alpha) /\tau }
\int_{\mathbb{R}^r} \rho(a; \tau) \, \e^{2 \pi i \sfQ(a) \tau + 2 \pi i \sfB(a,\alpha)} \dd a \right)
= \frac{\partial}{\partial \alpha_l} \left( \frac{1}{\sqrt{-\det A}} \frac{i}{(-i \tau)^{r/2}} \, \rho(\alpha;
-1/\tau) \right) \;.
\end{equation}
It follows that the expressions in the two brackets have to agree, up to an $\alpha$-independent expression.
Since $\rho(x)$ is an odd function, both brackets are odd as a function of $\alpha$,
and hence the $\alpha$-independent expression has to vanish.  This proves lemma \eqref{lemma-theta}.

Finally, we use the Poisson summation formula
\begin{equation}
\sum_{\nu \in \mathbb{Z}^r} f(\nu) = \sum_{\nu \in A^{-1} \mathbb{Z}^r} \tilde{f}(\nu) 
\end{equation}
to prove \eqref{transf-theta-tau}.
Here $\tilde{f}(\nu) = \int_{\mathbb{R}^r} f(a) \e^{2 \pi i \sfB(\nu, a)} \dd a$.
Using \eqref{theta-series} we infer
\begin{equation}
\vartheta_{a,b}(- 1/\tau) = \sum_{\nu \in \mathbb{Z}^r} f(\nu) 
\end{equation}
with 
\begin{equation}
f(\nu) = \rho(\nu + a; - 1/\tau) \, {\rm e}^{- 2 \pi i \sfQ(\nu + a)/\tau + 2 \pi i \sfB(\nu + a, b)} \;.
\end{equation}
Computing the associated $\tilde{f}(\nu)$ using \eqref{lemma-theta} yields
\begin{equation}
\tilde{f}(\nu) = \frac{i}{\sqrt{-\det A}} (-i \tau)^{r/2} \, \rho(\nu + b; \tau) \, {\rm e}^{2 \pi i \sfQ(\nu+b) \tau}
\, {\rm e}^{- 2 \pi i \sfB(a, \nu)} \;.
\end{equation}
Then, performing the sum $\sum_{\nu \in A^{-1} \mathbb{Z}^r} \tilde{f}(\nu)$ gives \eqref{transf-theta-tau}.

Now let us discuss the differentiation of $\vartheta(z; \tau)$ with respect to $a$. We begin by 
differentiating under the sum, 
\begin{equation}
\sum_{n \in \mathbb{Z}^r} \frac{\partial}{\partial a_l} \left(
\rho(n + a; \tau) \, {\rm e}^{ 2\pi i \,\tau \, \sfQ(n) + 2 \pi i \, \sfB(n,z)} \right) \;,
\end{equation}
where we recall that $ z = a \tau + b$ and 
$\rho = \rho^{c_1} - \rho^{c_2}$ with
\begin{equation}
\rho^c (n + a; \tau) = E \left(\frac{\sfB(c, n + a )}{\sqrt{- \sfQ(c)}} \, \sqrt{{\rm Im} \, \tau} \right) \;.
\end{equation}
We obtain
\begin{eqnarray}
&& \sum_{n \in \mathbb{Z}^r} \left[
E'\left(\frac{\sfB(c_1, n + a )}{\sqrt{- \sfQ(c_1)}} \, \sqrt{{\rm Im}  \, \tau} \right) \frac{\sqrt{{\rm Im} \, \tau} }{\sqrt{- \sfQ(c_1)}} \left(c_1 A\right)^l - 
E'\left(\frac{\sfB(c_2, n + a )}{\sqrt{- \sfQ(c_2)}} \, \sqrt{{\rm Im} \, \tau} \right) \frac{\sqrt{{\rm Im} \, \tau} }{\sqrt{- \sfQ(c_2)}} \left(c_2 A\right)^l \right.\nonumber\\
&& \left. \qquad + 2 \pi i \, \tau \, 
\rho (n + a; \tau) \, \left(n A \right)^l
\right]
\, {\rm e}^{ 2\pi i \,\tau \, \sfQ(n) + 2 \pi i \, \sfB(n,z)} \;.
\label{summ-3}
\end{eqnarray}
Let us consider the first summand,
\begin{eqnarray}
&&\left|E'\left(\frac{\sfB(c_1, n + a )}{\sqrt{- \sfQ(c_1)}} \, \sqrt{{\rm Im} \, \tau} \right) 
\, {\rm e}^{ 2\pi i \,\tau \, \sfQ(n) + 2 \pi i \, \sfB(n,z)} \right| \nonumber \\ && \qquad = 
\left|{\rm e}^{\pi \, \sfB^2(c_1, n + a ) \, {\rm Im} \, \tau/\sfQ(c_1) } 
\, {\rm e}^{ 2\pi i \,\tau \, \sfQ(n) + 2 \pi i \, \sfB(n,z)} \right| \nonumber\\
&& \qquad =  \left|{\rm e}^{\pi \, \sfB^2(c_1, n + a ) \, {\rm Im} \, \tau/\sfQ(c_1) } 
\, {\rm e}^{ 2\pi i \,\tau \, \sfQ(n + a) + 2 \pi i \, \sfB(n,b) - 2\pi i \,\tau \, \sfQ(a)} \right| 
\nonumber\\
&& \qquad = {\rm e}^{2\pi {\rm Im \, \tau} \, \sfQ(a)} \, {\rm e}^{- 2 \pi \left[\sfQ(n + a) - \sfB^2(c_1, n+a)/(2 \sfQ(c_1)) \right]
{\rm Im} \, \tau } \;.
\end{eqnarray}
By Lemma 2.5 of \cite{Zwegers:2008}, the series
\begin{equation}
\sum_{\nu \in a + \mathbb{Z}^r}  {\rm e}^{- 2 \pi \left[\sfQ(\nu) - \sfB^2(c_1, \nu)/(2 \sfQ(c_1) )\right]
{\rm Im} \, \tau } 
\end{equation}
converges, and for ${\rm Im} \, \tau \geq \varepsilon > 0$ the series converges uniformly, since
\begin{equation}
{\rm e}^{- 2 \pi \left[\sfQ(\nu) - \sfB^2(c_1, \nu)/(2 \sfQ(c_1) )\right]
{\rm Im} \,  \tau } 
\leq
{\rm e}^{- 2 \pi \left[\sfQ(\nu) - \sfB^2(c_1, \nu)/(2 \sfQ(c_1) )\right]
\varepsilon} \;.
\end{equation}
Now consider the last summand of \eqref{summ-3},
\begin{eqnarray}
&& \left| \rho (n+ a, \tau) \left(n A \right)^l
\, {\rm e}^{ 2\pi i \,\tau \, \sfQ(n) + 2 \pi i \, \sfB(n,z)} 
\right| \leq \left| \left(n A \right)^l \right| {\rm e}^{2\pi {\rm Im \, \tau} \, \sfQ(a)}
\,
\left[  {\rm e}^{- 2 \pi \sfQ^+ (n + a) 
{\rm Im} \tau}
 \right. \nonumber\\
&& \left. + \,  {\rm e}^{- 2 \pi \left[\sfQ(n + a) - \sfB^2(c_2, n+a)/(2 \sfQ(c_2)) \right] {\rm Im} \, \tau } + {\rm e}^{- 2 \pi \left[\sfQ(n + a) - \sfB^2(c_1, n+a)/(2 \sfQ(c_1)) \right]
{\rm Im} \, \tau }  \right] \;,
\label{summand-3}
\end{eqnarray}
where $\sfQ^+$ denotes the positive definite quadratic form introduced in lemma 2.6 of \cite{Zwegers:2008}.
Given a positive definite quadratic form $\tilde \sfQ : \mathbb{R}^r \rightarrow \mathbb{R}$, there exists $\delta >0$
with $\delta \in \mathbb{R}$ such that $2 {\tilde \sfQ} (\nu) \geq \delta \sum_i (\nu_i)^2 $, and hence
\begin{equation}
{\rm e}^{- 2 \pi {\tilde \sfQ} (\nu) {\rm Im} \, \tau } \leq {\rm e}^{- \pi \delta \, {\rm Im} \, \tau \, \sum_{i=1}^r 
(\nu_i)^2} \;.
\end{equation}
Thus we obtain
\begin{eqnarray}
\left| \left(n A \right)^l \right| {\rm e}^{- 2 \pi {\tilde \sfQ} (n+ a) {\rm Im} \, \tau } \leq 
\left| \left(n A \right)^l \right| 
{\rm e}^{- \pi \delta \, {\rm Im} \,  \tau \, \sum_{i=1}^r 
((n + a)_i)^2} \;.
\end{eqnarray}
Now we consider the terms involving $n_l$, 
\begin{eqnarray}
| n_l | {\rm e}^{- \pi \delta \, {\rm Im} \, \tau \, 
((n + a)_l)^2} \;.
\end{eqnarray}
Taking $n_l >> 1$ and assuming $a_l \geq 0$, for simplicity, we get
\begin{eqnarray}
n_l \, {\rm e}^{- \pi \delta \, {\rm Im} \, \tau \, 
((n + a)_l)^2} \leq  n_l \, {\rm e}^{- \pi \delta \, {\rm Im} \, \tau \, 
n_l^2} \leq  n_l \, {\rm e}^{- \pi \delta \, {\rm Im} \, \tau \, 
n_l} \;.
\end{eqnarray}
Using that the sum
\begin{equation}
\sum_{n \geq 1} n \, e^{- n t }
\end{equation}
is uniformly convergent for $t \geq \varepsilon > 0$, we conclude that the series obtained by summing over
\eqref{summand-3} is uniformly convergent for ${\rm Im} \tau \geq \varepsilon > 0$, and so is \eqref{summ-3}.
Hence it follows that
\begin{equation}
\sum_{n \in \mathbb{Z}^r} \frac{\partial}{\partial a_l} \left(
\rho(n + a; \tau) \, {\rm e}^{ 2\pi i \,\tau \, \sfQ(n) + 2 \pi i \, \sfB(n,z)} \right) = 
\frac{\partial }{\partial a_l} \vartheta(z; \tau) \;.
\end{equation}

\section{An example of an indefinite theta function} \label{thetaexample}

In this appendix we consider an explicit example \cite{GottZag1996} of an indefinite theta function.  In doing so we explicitly show how 
indefinite theta functions differ from ordinary theta functions, and how the indefinite directions are dealt with. In this example the weight function $\rho$ is taken to be the difference of two sign functions. In this case Zweger's ``wedges" act as a projector onto a specific sublattice. We consider the indefinite theta function given by  \cite{GottZag1996}
\begin{equation}
\sum_{x \in \Gamma^{1,1}} \ \rho (x + \alpha) \ \e^{2 \pi i \sfQ(x) \tau + 2 \pi i \sfB (x,\gamma)} \ ,
\end{equation}
defined over the indefinite lattice $\Gamma^{1,1}$. We write the vector $\gamma$ as $\gamma = \alpha \tau + \beta$. The other vectors are explicitly
\begin{equation}
x = (m,n) \ , \qquad \gamma=(\gamma_1 , \gamma_2) \ , \qquad \alpha=(\alpha_1 , \alpha_2) \ , \qquad \beta = (\beta_1 , \beta_2) \ .
\end{equation}
For simplicity we will focus on the case where both $c_1$ and $c_2$ are chosen in such a way that $\sfQ (c_i) = 0$ for $i=1,2$. In this case the weight function $\rho = \rho^{c_1} - \rho^{c_2}$ simply plays the role of a projector and 
\begin{equation}
\rho^{c_i} (x + \alpha) = \sign \, \sfB \left( x + \alpha ; c_i \right) \ .
\end{equation}
The condition $\sfQ (c) = 0$ has four solutions: $(0, \pm 1)$ and $(\pm 1 , 0)$. From these we can construct two chambers where $\sfQ$ is negative: $\cS_1$ delimited by $c_1 = (0,+1)$ and $c_2 = (-1,0)$ and $\cS_2$ delimited by $c_1 = (0,-1)$ and $c_2 = (1,0)$. Furthermore $\tau$ and $\gamma$ have to lie within the domain \cite{GottZag1996}
\begin{equation}
\cD (c) = \{  0 < \im (c \cdot \gamma) < \im \tau  \} \ .
\end{equation} 
The domain $\cD (c)$  can be written explicitly for the four vectors such that $\sfQ (c)=0$ as follows,
\begin{eqnarray}
\cD (c=(0,+1)) &=& \{ 0 < \alpha_1 < 1 \} \;, \nonumber\\
\cD (c=(0,-1)) &=& \{ -1 < \alpha_1 < 0 \} \;, \nonumber\\
\cD (c=(+1,0)) &=& \{ 0 < \alpha_2 < 1 \} \;, \nonumber\\
\cD (c=(-1,0)) &=& \{ -1 < \alpha_2 < 0 \} \;.
\end{eqnarray}

We have now two possibilities.
In the first case we pick the domain $\cS_1$. Then $c_1 = (0,+1)$ and $c_2 = (-1,0)$. Therefore we have to work with the condition
\begin{equation} \label{conditions1}
 \{ 0 < \alpha_1 < 1 \} \cap  \{ -1 < \alpha_2 < 0 \} \;.
\end{equation}
In particular
\begin{eqnarray}
\rho (x+a ; \tau) &=& \sign \, \sfB(x+\alpha , c_1) - \sign \, \sfB (x+\alpha , c_2) \nonumber\\
 &=& \sign (m+\alpha_1) + \sign (n+\alpha_2) = \left\{ \begin{matrix} 
 +2 & \text{if} & m>0 \, , n> 0 \\
 -2 & \text{if} & m<0 \, , n< 0 \\
 0 & \text{if} & m>0 \, , n< 0 \\
 0 & \text{if} & m<0 \, , n> 0 \\ 
 \end{matrix} \right.
 \label{contr1}
\end{eqnarray}
The border values have to be treated separately:
\begin{eqnarray}
m&=&0 \Longrightarrow \sign(\alpha_1) + \sign (n+\alpha_2) = +1 + \sign (n+\alpha_2) \neq 0 \iff n>0 \nonumber\\
n&=&0  \Longrightarrow \sign(m+\alpha_1) + \sign (\alpha_2) = \sign (m+\alpha_1) - 1 \neq 0 \iff m<0
\end{eqnarray}
Therefore the sum of the theta function is only in $m \ge 0 , n>0$ and $m < 0 , n \le 0$. Therefore
\begin{eqnarray}
&& \sum_{x= (m,n) \in \Gamma^{1,1}} \rho (x+\alpha)  \, q^{mn} \ \e^{m ( 2 \pi i \gamma_2) + n (2 \pi i \gamma_1)}  \nonumber \\
&& \qquad \qquad 
 = 2 \sum_{\stackrel{m \ge 0}{n>0}} q^{mn} \ \e^{m v+n u} - 2 \sum_{\stackrel{m >0}{n \ge 0}} q^{mn} \ \e^{-m v - n u}
 \;,
\end{eqnarray}
where we have set $u= 2 \pi i \gamma_1$ and $v = 2 \pi i \gamma_2$, and where $q = {\rm exp} [2 \pi i \tau]$.

In the second case we choose the domain $\cS_2$ and pick the two vectors as $c_1 = (0,-1)$ and $c_2 = (1,0)$. This means
\begin{equation}
 \{ -1 < \alpha_1 < 0 \} \cap  \{ 0 < \alpha_2 < 1 \} \;,
\end{equation}
which implies
\begin{eqnarray}
\rho (x+a ; \tau) &=& \sign \sfB(x+\alpha , c_1) - \sign \sfB (x+\alpha , c_2) \nonumber\\
 &=& - \sign (m+\alpha_1) - \sign (n+\alpha_2) = \left\{ \begin{matrix} 
 -2 & \text{if} & m>0 \, , n> 0 \\
 +2 & \text{if} & m<0 \, , n< 0 \\
 0 & \text{if} & m>0 \, , n< 0 \\
 0 & \text{if} & m<0 \, , n> 0 \\ 
 \end{matrix} \right.
\end{eqnarray}
Now the border cases are
\begin{eqnarray}
m&=&0 \Longrightarrow \sign(\alpha_1) - \sign (n+\alpha_2) = +1 - \sign (n+\alpha_2) \neq 0 \iff n<0 \nonumber\\
n&=&0  \Longrightarrow \sign(m+\alpha_1) + \sign (\alpha_2) = - \sign (m+\alpha_1) - 1 \neq 0 \iff m>0
\end{eqnarray}
The sum is now only over $m \le 0 , n<0$ and $m > 0 ,n \ge 0$. This means
\begin{eqnarray}
\sum_{m,n} && \rho(x + \alpha)  \, q^{mn} \ \e^{m ( 2 \pi i \gamma_2) + n (2 \pi i \gamma_1)} \nonumber\\
&=&  - 2 \sum_{\stackrel{m >0}{n \ge 0}} q^{mn} \ \e^{m v + n u} +
2 \sum_{\stackrel{m \ge 0}{n>0}} q^{mn} \ \e^{-m v-n u} \;.
\end{eqnarray}

Note that in both cases the weight function $\rho$ acts as a regulator projecting out certain lattice points, among which those that would have given an exponentially growing contribution. The situation where $\rho$ is the difference between two error functions (\ref{rho-x}) can be treated similarly. Indeed as pointed out in (\ref{rho-x2}) the error function can be written as   
\begin{equation} \label{errappC}
E(x) =  {\rm sgn}(x)\;\left( 1 - \beta(x^2) \right) =  {\rm sgn}(x)  -  {\rm sgn}(x) \beta(x^2)     \ .
\end{equation}
The first term in the sum behaves precisely as we have explained in the above text. The second term has a different role. Note that the second term is precisely the combination which we called $\tilde{\rho}$ in the appendix \ref{app-ind-theta-fun}, cf. below \eqref{theta-series}.  As discussed there, it follows from (\ref{series-Qc}) that the combination of $\tilde{\rho}$ with the exponential of an indefinite quadratic form is always bounded by a damped positive definite quadratic form. This was used to bound the whole series to prove its convergence. This means that in the function $\rho = \rho^{c_1} - \rho^{c_2}$, the competition between the sign terms in (\ref{errappC}) acts as a projector, while the terms proportional to $\beta$
sum over all the lattice points, but suppress their contribution exponentially via the error function.


\providecommand{\href}[2]{#2}\begingroup\raggedright\endgroup

\end{document}